TALLINN UNIVERSITY OF TECHNOLOGY
School of Information Technologies

Teona Gelashvili 184725IVGM

# GOING PAPERLESS:
# MAIN CHALLENGES IN EDRMS IMPLEMENTATION.
# CASE OF GEORGIA

# PEAMISED TAKISTUSED EDRMSI JUURUTAMISEL
# PABERIVABALE ASJAAJAMISELE ÜLEMINEKUL.
# GRUUSIA JUHTUMIUURING

Master's Thesis

Supervisor: Ingrid Pappel
Prof. Dr.
Co-Supervisor: Valentyna Tsap
MSc

# Author's declaration of originality

I hereby certify that I am the sole author of this thesis. All the used materials, references to the literature and the work of others have been referred to. This thesis has not been presented for examination anywhere else.

Author: Teona Gelashvili

07.05.2020

# Abstract

National governments are eager to incorporate information and communication technologies in their administrative agenda. Even though transformation from traditional to digital methods seem attractive, there are obstacles which come along with this change. Since countries are continuously implementing ICT-based technologies, the need for investigating the factors hindering their adoption becomes crucial.

The main objective of this study is to inquire Electronic Documents and Records Management Systems (EDRMS) in the context of eGovernment. The centre of the investigation is how EDRMS could raise efficiency in public service delivery. Within the scope of this research, the associated challenges of the EDRMS implementation, possible ways to tackle those challenges and achieved benefits were analysed. For thorough examination, exploratory case study method was chosen, and the case of Georgia was investigated. For drawing conclusions and finding answers to the imposed questions qualitative approach was selected. Different ICT adoption theories and case study examples were analysed, among which the Estonian case was taken as a successful model. Empirical data collection was also an integral part of the study during which interviews with state officials, as well as experts, were conducted, and the questionnaire among civil servants was distributed.

As a result of the research, the main aspects of the EDRMS implementation were evaluated, and the prospect of the possible future study was imposed.

This thesis is written in English and is 65 pages long, including 6 chapters, 11 figures and 3 tables.

# List of abbreviations and terms

| | |
|---|---|
| BPM | *Business Process Management* |
| DIP | *Document Image Processing systems* |
| DOI | *Diffusion of Innovations* |
| EDMS | *Electronic Documents Management Systems* |
| EDRMS | *Electronic Documents and Records Management Systems* |
| EMIS | *Electronic Management Information System* |
| ERMS | *Electronic Records Management Systems* |
| G2B | *Government to Business* |
| G2C | *Government to Citizen* |
| G2G | *Government to Government* |
| ICT | *Information and Communication Technology* |
| IT | *Information Technology* |
| LEPL | *Legal Entity of Public Law* |
| OGP | *Open Government Partnership* |
| PRONI | *The Public Record Office of Northern Ireland* |
| RO | *Research Objective* |
| ROI | *Return of Investment* |
| RQ | *Research Question* |
| TAM | *Technology Acceptance Model* |
| TNA | *United Kingdom's National Archives* |
| TPB | *Theory of Planned Behaviour* |
| TRA | *Theory of Reasoned Action* |
| UTAUT | *Unified Theory of Acceptance and Use of Technology* |
| UX | *User Experience* |

# Table of contents

# List of tables

# List of Figures

# 1 Introduction

## 1.1 Overview of Research

e-Governance and e-Government have become prevalent terms among countries across the world. Scholars have developed various interpretations of how to define them. According to Bose and Rashel the main principle of e-Governance is the delivery of the governmental services through the means of Information and Communication Technology (ICT), which also incorporates increased communication between Government to Government (G2G), Government to Cityzen (G2C) and Government to Business (G2B) [1], [2] Heeks describes e-Government as "The use of IT by public sector organizations." [3] Therefore, efficient service delivery, increased interaction of government to citizens and businesses and the application of ICT are core principles of e-Governance [4], [5].

In the mid-1990s, it was the government of the USA, Great Britain, and other technologically advanced Western countries, starting to use the internet for delivering governmental services. Nowadays, the list is much longer, including many other countries [6]. IT is now an integral part of the provision of governmental services and is a key and vital resource. It is, first and foremost, an enabling technology that can be embraced by the workforce to improve organizational efficiency and effectiveness. To understand whether "Technologies help to govern and to be governed" [7], it is vital to discuss not only technological but also social aspects of it.

The development of the IT infrastructure has been considered to be a significant factor for successful e-service delivery during the past few decades. A strong public-private partnership, as well as interoperability between systems, can boost the e-service development [8]. However, failed projects made it clear that people's interaction with technology is not only rational, but it also includes emotional factors [9], [10], [11]. As we will also see from the theories discussed in the upcoming chapters below, it is essential to understand not only how the organization, as a whole, reacts towards the implementation of the new information technology but also, what is the reaction at the individual (staff) level [12]. Despite the complexity of the matter, there are still countries with successful EDRMS implementation examples. For instance, Estonia known as digital state/tech-savvy country/e-state has introduced paperless management since the beginning of 2000 and has achievements based on the proposed implementation methodology [13], [14] which involved both human and information system aspects.

Public organizations are aiming to sustain with the ongoing trend of the IT system development and come up with the innovative ways of altering traditional service delivery not only within the organizations but externally as well – for the citizens [15], [16]. It can have a tremendous impact and even change the way an organization undertakes tasks within the prevailing organizational culture. One of the examples of it is the implementation of the Electronic Documents and Records Management Systems (hereafter EDRMS). However, taking into consideration the fact that EDRMS implementation is closely linked to the social factor, as it is not only technological innovation, there is also a possibility of numerous obstructions [17].

February 21st 2012, the Government of Georgia issued the Decree on the establishment of the minimal standards of the usage of EDRMS in public organizations across the country [18]. According to the Decree, all public organizations were obliged to meet the standards latest by June 1st, 2012. The programs public organizations use varies. Nevertheless, the core principle they all share is to improve the official correspondence and document exchange between public organizations, manage tasks efficiently, and reduce administrative burden. This change also meant to facilitate the development of e-Governance in the country. However, as Adam says: "Implementing EDRMS is not just about technology—that is the easy part! It is more about people, organizations, organizational culture, cultural change, and good, strong, yet flexible project management." [15].

The core principle of this work is to gather empirical as well as secondary data, study EDRMS implementation experiences shared by different countries, look at them through the lens of various technology adoption theories, learn from the successful Estonian model and based on this accumulated knowledge, analyse the challenges of EDRMS implementation in Georgian governmental institutions. These findings hopefully, will be beneficial for future studies. As emphasised before, a paperless approach through the usage of EDRMS has been widely applied in the Estonian public sector. Although there have been many challenges and obstacles, Estonia managed to tackle and solve them over the years and raised efficiency [13]. Now, the country has been rewarded in a digital document exchange level with more than 90% between public agencies [19]. Estonia will be discussed as a model of successful transformation which Georgian public sector aspired to achieve.

## 1.2 Research Motivation

Public organizations are inclined to the idea of implementing an Electronic Document and Records Management System (EDRMS) and alter their traditional way of working. With the hope that implementation can resolve complex business problems, deliver real competitive advantage, and transform organizations, governments are investing resources in adopting IT solutions. The very first motivation is to discover how public sector efficiency is increased due to these technological changes, as based on the studies shifting to paperless management has caused the significant organizational transformation, and the outcomes are worth learning.

Another essential aspect that is concerning this change is the human resistance factor. There is a plethora of research made to investigate possible pitfalls in terms of infrastructure, lack of funds, or possible legal issues. Nevertheless, the topic or understanding of how employees react to such changes is still not significantly covered. If we are talking about the transformation and not only digitations of the workflow, possible pitfalls governments encounter should be clearly defined when dealing with the implementation and further utilization of EDRMS.

I am especially curious to discover this topic further since I was working in one of the public organization for almost three years. The entity I was working for is under the supervision of the Ministry of Culture of Georgia, and we were able to communicate with other public institutions regarding most of the topics via EDRMS. Looking back now, I realize how much more beneficial and efficient my workday could have been if I was able to use all the benefits this system offered. Remembering how I used to fulfil my everyday tasks is allowing me to understand what issues we have been encountering, and I am willing to find solutions by researching them thoroughly now.

The final factor is that e-Governance in Georgia is the topic which has not been studied by the researchers comprehensively yet. During the whole process of working on this thesis, I had a hard time finding records, data, information, which would ease my research. Hopefully, my work will be useful for future researchers who are interested in investigating this emerging yet already quite a growing field and alongside have data about the country.

## 1.3 Research Questions and Objectives

As mentioned earlier, the Government of Georgia continuously highlights the importance of e-Governance; however, there is not much research done in this field and the processes analyzed. This study is built around the examining process of introducing EDRMS in Georgian governmental institutions on the premise that achieving employees' acceptance of this new technology is vital to its successful implementation. The transformation from the traditional to the electronic document management system organizations were challenged to combat various difficulties emerging in the course. Finding the way to tackle them and achieve organization-wide benefits is the goal of a successful transition. To investigate the implementation process further (at the national and local government level), an overall question was formed: *how EDRMS could raise efficiency in the public service delivery* and in order to answer that, the three research questions below were drafted, which were guiding through the entire study:

- *RQ1: What were the EDRMS implementation challenges in the Georgian governmental institutions?*

In order to become aware of all the aspects affecting the EDRMS implementation at a considerable level, systematic examination of the case will be carried out. Answering this question will allow a thorough analysis of all the technical and social impediments emerging in the process of carrying out this significant transformation and the national and local government level.

- *RQ2: How did Georgian government entities tackle those challenges?*

Presented second research question derives from the first one and aims to investigate all the measures which were taken by the Georgia governmental institutions to overcome the difficulties. Practices executed by the national and local governments will be highlighted to paint a full picture of the situation.

- *RQ3: What benefits have been achieved through EDMRS implementation in the Georgian government entities?*

To conclude, achieved benefits of shifting to the paperless management from both government institutions and citizens perspective will be assessed. Evidence of how this change is reflected in the communication between the public institutions, their mundane activities and service delivery will be collected and analysed.

Based on the scope of this study, the qualitative method was adopted to gathering empirical data, to provide comprehensive answers to the imposed research questions. For establishing the agenda of the future analysis, below Research Objectives were determined:

**RO1:** Conduct a thorough literature review to analyse EDRMS implementation-related issues identified in different countries, becoming familiar with various technology acceptance and change management theories within the frame of the ICT adaptation factors among employees

**RO2:** Inspecting practical EDRMS implementation experiences among Georgian and Estonian national and local government entities (conducting interviews with state officials and experts)

**RO3:** Examine employees' opinions about the EDMRS usage in their daily activities, also what is the range of the daily assignments they accomplish utilizing the system (qualitative method of allotting online survey was used here)

**RO4:** Detect the policies undertaken by the public institutions at both - national and local government levels, in order to combat EDRMS implementation challenges; in addition, to distinguish achieved benefits (this topic was monitored form the government officials as well as civil servants viewpoints)

## 1.4    Research Design and Methodology

After formulating the presented research questions, a suitable methodology was adopted for a rigorous investigation of the matter. Analyzing EDRMS usage in Georgian public organizations requires thorough observational shreds of evidence to be gathered, and qualitative research methods suite the aim of this study the most.

Three major traditions developed by Plato, Aristotle, and Greek humanities were prevalent in the Western intellectual world over the centuries. According to them, there is one absolute truth aimed to discover. Another approach which continued their concept is the one developed by Emile Durkheim, according to which researcher's aim is "to use the knowledge obtained from observation to discover the reality and find the causes of the social phenomenon" [20]. However, as Kincheloe and Berry emphasized, the unilateral perspective will not be enough to cover the complexity of the subject matter [21]. The multilateral perspective is guiding through this research and will be discussed more in details below.

Taking into consideration the fact that the researcher is the frontrunner and establishes the framework of the project, their role requires further courtesy. Willis highlights that researchers should focus on the situated and contextual understanding (observation, action and interaction in the specific context as highlighted by micro-sociologists) instead of searching for the "truth". To be open to multiple perspectives and sources of data, take foundational, holistic and not technique, and atomic approach is crucial. Therefore, something Willis describes the qualitative researcher as "a loosely scheduled traveller than the traveller, who makes detailed plans, with all the stops and routes set in advance" which is appropriate for the setting of this study [22].

It is the nature of qualitative research the findings to emerge during the examination of the collected data [23]. Grounded Theory, which is flexible in terms of providing the opportunity of changing initial study questions and expound methodology alongside with evaluation of the acquired data, has been chosen as a paradigm of this study [22]. Per this framework, whole work can be divided into three major parts: a collection of the raw data, develop a tentative theory based on that data and then collect more data to justify the opinion above.

One of the research methods of this study is an inductive case study – "an examination of a specific phenomenon" [22]. According to Yin, "a case study is an empirical method that investigates a contemporary phenomenon in depth and within its real-world context." [24] Four significant reasons are justifying the selection of this method. Firstly, it provides the possibility of explaining complex presumed causal links in real-world interventions. Secondly, it describes the response and real-world context in which it occurred. Thirdly, it can illustrate specific topics within the evaluation. Finally, case study research helps enlighten those situations in which the observation has no clear, single set of outcomes [24].

There are four sources of data used in this case study, such as documentation (administrative documents, formal studies and evaluations), archival records (including different organizational records, publicly-accessible files), semi-structured interviews and survey results.

According to Weber's concept of "Verstehen"/understanding, we need to "get into the heads" of the people we study, which means acquiring a deeper understanding of the individuals we are interviewing [20]. Therefore, the importance of interviews was highlighted in this study. They gave opportunity investigation to be pointed and rapt around the research subject. In addition to that, feminists, while disregarding "masculinist approach", emphasized how important it is for the researcher not to be detached from the research subject. In preference of judging from the omniscient position, imposing categories and variables on the data ("objectifiers"), feminists believed that researchers should have become research participants instead of having subjects and objects [20].

Tactics discussed above, starting with the concept of the researcher, research object, the possibility of finding the truth and the role of the researcher, were the main principles when preparing for the interviews and each method in general. In order to achieve a comprehensive insight into the subject, it is vital to study it in its setting. It was the main criteria for choosing the respondents and exploring the case with the different stakeholder's perspective. Five interviewees are participating in this research. Three of them are from Georgia, and the remaining two shared Estonian experience in terms of EDRMS implementation. Georgian interviewees include Minister of the Education and Science of Georgia during 2009-2012, Head of the Document Management of the Ministry of Culture and Monument Protection of Georgia and the Chairman of the Lagodekhi

Municipality City Council. Their involvement in the development of the EDRMS in the country determined their contribution to this study.

Dimitri Shashkini was a Minister of the Education and Science of Georgia during 2009-2012 when the first official government decree about the standardization of the EDRMS was published. Within the framework of the Ministry, he was leading the projects in collaboration with other public entities aiming to achieve improved communication between the organizations and better service delivery.

Ekaterine Gurgenidze, Head of the Document Management of the Ministry of Culture and Monument Protection of Georgia, has been working in the Ministry since 2010. She was actively engaged in the implementation of the EDRMS in 2012, and her direct duty was to choose the system which would suit the entity the most. Therefore, she gave excellent insight and described all the struggles they have been through during the implementation.

The Chairman of the Lagodekhi Municipality City Council Karlo Jamburia is also a person who initiated acceleration of the EDRMS implementation at the local government level. Before becoming the Chairman, he was Governor of the municipality for a few years, and he implemented the system there as well. His input was valuable not only for a better understanding of the situation at the local government level but also to recognise the role of the leadership when initiating such a significant change.

As mentioned in the introduction, the Estonian case will be discussed in this work as the model of successful practice. However, only investigating secondary data and inquiring documents would not comprehensively fulfil the scope of this study. As discussed above, an in-depth investigation of the case is the principal factor to the thorough analysis. It is the reason why two Estonian experts were interviewed. Liivi Karpištšenko, the prominent professional in her field, provided a unique contribution by disclosing details of the EDRMS implementation in Estonian public sector and bringing up possible shortcomings they are facing now. Her background in ICT and experience of working in document management systems over the years in different public entities was very beneficial for this research. Currently, she is employed in the Ministry of Economic Affairs and Communications. In order to have a wholesome understanding of the case, the consultant's insight into this case is valuable. Mihkel Lauk, PwC Advisors IT Senior Consultant, provided input regarding Estonian experience and discussed some common challenges, pitfalls and shortcomings which usually come along with the EDRMS implementation process.

All of the interviews were recorded and transcribed. For three interviews conducted in Georgian, transcription as well as translation from Georgian into English was done manually by the author of this thesis. For the thematic analysis and drawing mind and project maps, NVIVO was the tool to use. Through coding and creating various notes in this software, it was convenient to identify, analyse and interpret patterns ("themes") and start concluding.

Furthermore, it was essential to identify the EDRMS attitudes of the people working in the public sector. The survey method was handy and allowed examining the more significant part of the population, gathering data, and most importantly, it was possible to collect information remotely. Respondents of the survey were civil servants from various public organisations using several systems. The survey consisted of multiple-choice as well as open-ended questions which aimed to investigate their views towards the usage of EDRMS and correlation with different variables. Survey Monkey was a tool used for constructing the questions and distributing. SPSS is the tool used for generating descriptive statistics, analysing, identifying correlations, and analysing the results. The questionnaire and charts describing the results are appendices at the end of this research.

## 2. Literature Review

This chapter intends to provide a comprehensive overview of the available theories addressing ICT implementation as well as country-specific case studies. In the first section, all the major theories which address various aspects of the technology implementation and the dependence of the users will be discussed. Even though they have been developed in different periods, there are common factors highlighted by each of the authors. The core principles of the theories, underlined concepts and linkages are listed per the time frame they were developed. At the end of this chapter, the table lists vital technology acceptance factors identified in each of the below-discussed theory.

The second section examines and briefly summarises the experiences of implementing EDRMS in different countries. It portrays the challenges and obstacles national governments went through, what benefits were achieved and what was the reaction from the staff members. After finishing this research, comparing the findings to those effects identified by countries with different settings will be beneficial for the reliability of this study. Like in the case of the previous section, this one will also have a table with the list of the countries, the problems they identified and lessons they learned.

EDRMS usage is radically altering the way public sector employees used to work with physical documents and records. Therefore, there is this significant transformation of organizational systems and processes, which is closely linked to the human factor, and thus, the need for cultural change is vital. The third section of this chapter will be dedicated to understanding the importance of managing cultural change.

## 2.1 Theories about ICT Implementation and Adoption

Scholars have been curious about investigating various approaches, which attempt to define aspects of ICT implementation and adaptation. One of the very first theories to discuss here is the Theory of Reasoned Action (TRA) developed by Martin Fishbein and Icek Ajzen in 1967, which stems from social psychology [25]. The core principle of the TRA is that it tries to outline the existing connection of people's subjective norms "the person's perception that most people who are important to him think he should or should not perform the behaviour in question" towards their beliefs and attitude [26]. It mainly focuses on identifying what people's behaviours are depending on their pre-existing attitudes.

Slightly more than two decades later, Cooper and Zmud shifter the focus from the personal perceptions and developed IT Implementation Process Model [27]. Something that dramatically differentiates this model from TRA is that here exact six stages of ICT adaptation is outlined at the organizational level. Unlike TRA link is not to the social science but works related to organizational change and innovation [25]. The six stages model begins from the initial stage of implementation, which is about understanding the organization itself and finishes with the adoption and efficient consumption of the technology by the staff members [27]. The stages mentioned are: "initiation, organizational adoption, adaptation, acceptance and adoption, routinization, and infusion." [25].

Diffusion of Innovations (DOI) Theory offered a broader approach to the study of the subject. Two integral factors this theory focuses on are the distribution of new ideas within the organization and the importance of communication and leadership on their acceptance [28]. Later, Moore and Benbasat utilized DOI to develop "An instrument designed to measure the various perceptions that an individual may have of adopting an information technology (IT) innovation" [29]. The core principle of this instrument was to understand the primary factors of why individuals decide to accept the technology and how it spreads across the organization at the later stage.

The need for understanding and predicting why and when users choose to adopt and use technology was also a fundamental principle in the subsequent Technology Acceptance Model (TAM) developed in 1989 by Fred Davis [30], [9] with two ultimate elements of system use [25], [31], [26]. These two theoretical constructs are perceived usefulness "The degree to which a person believes that using a particular system would enhance his or her job performance", and perceived ease of use "The degree to which a person believes that using a particular system would be free of effort" [30]. Even though the theory is not recent, the hypothesis of numerous case studies is built on its framework.

Theory of Planned Behavior (TPB) by Icek Ajzen was developed in 1991 and aimed to improve the predictive aspect of his previous TRA theory and put emphasis on cognitive self-regulation in the sense of perceived behavioural control [32]. Like it was in the TRA, personal attitudes are also integral here in terms of shaping human behaviour. Also, two more determinants are influencing the adoption, such as social pressures and a sense of control [33]. DeLone and McLean looked up the existing studies within the context of the TPB theory and presented the list of the categories measuring Success of Information Systems: "System quality, information quality, use, user satisfaction, individual impact, and organizational impact [34]."

The last theory to discuss is the Unified Theory of Acceptance and Use of Technology (UTAUT) by Venkatesh, Morris and Davis. Just like in the TAM, UTAUT as a continuation of it also attempts to understand the reasons affecting users when adopting the technology but also adds another determinant – the role of Managers [35], [36]. There are four vital moderating variables: "Gender, age, experience, and voluntariness of use" and central concepts: "Performance expectancy, effort expectancy, social influence, and facilitating conditions." influencing user acceptance and user behaviour [25], [10], [37].

In conclusion, all the major theories discussed above are concentrated on examining factors which determine the reasons why users decide to accept new technology. Regardless of what is the integral determinant, each of them separately is focusing on. The most cited theories of previous publications showed that the theory of the research area focused on the acceptance and adoption of technology. Short description of the main factors each theory focuses on is in the table below.

Table 1. Main Factors Affecting ICT Acceptance. Source: Author

| Identified factors | Source |
| --- | --- |
| Personal attitudes and perceptions | [25], [26], [28], [ [29]] |
| Organizational change | [27] |
| Role of leadership | [28], [29], [35] |
| Perceived usefulness and perceived ease of use | [9], [25], [30], [32], [33] |
| Social pressures and a sense of control | [32], [33] |
| The importance of expectancies | [35] |

## 2.2 Country Specific Case Study Examples

As mentioned earlier, the implementation of the EDRMS became a priority for many countries. Scholars have been investigating different aspects of their objectives, usage, benefits, obstacles, and lessons to be learned [38]. Some of the country-specific examples are below.

The first example to mention is the implementation of EDRMS at the Public Record Office of Northern Ireland (PRONI) in 2003. The author of the study examines all the major benefits achieved upon implementation and as he concludes the most crucial one was improvement within the internal records management procedures, which made a substantial difference at the level of the Northern Ireland Civil Service [39].

In early 2001, the Estates Department of the British Library introduced system called TRIM Captura. The goal was to share information, support document control and make information easier to find. The study shared experience of implementation and pointed out the following lessons learned from it: user-friendliness is crucial, metadata entry should be as automated as possible and that they should promote records management principles before introducing a system [40].

The fact that EDRMS was made obligatory within the public sector was a good foundation for the whole implementation process in Estonia. Respective acts were drafted in order to facilitate the process and the legal framework to be supportive of this transformation. An important aspect was practising collaboration among municipalities in order to harmonise public services [14].

Together with the achieved benefits, scholars also identify particular challenges which come along with the system implementation. It has been demonstrated in the study investigating EDRMS implementation in Australian public sector in 2009. The study revealed that critical hindrances occurred due to the resistance from the staff members due to their attachment to the traditional paper-based system [41]. Also, there was a difficulty encouraging them to share information as it has not been actively practised before. These obstacles have been further examined in another study, and as the author concludes it is difficult to have a clear endpoint since the implementation of EDRMS is directly linked to the cultural change and thus does not fit into a simple scratch of a strategy [42]. Taking into consideration the fact that change is inevitable, there is a need organisation to manage it upfront. As IS Manager, City of Charles Sturt, said: "A records system is an octopus - tentacles everywhere." [42]

Obstacles related to the EDRMS implementation in the New Zealand public sector organizations, as well as achieved benefits, have been examined by Yin. The benefits, perceived by the users are: new system helping to improve the overall information quality and efficiency, centralized repository with improved version-control, enhanced the process as it provides a single and safe repository for information storing. As for the drawbacks, the essential elements identified by the researcher include lack of training, improper support from the senior management's side, lack of proper records sharing culture pre-implementation [43].

Similar design of identifying obstacles and achieved benefits connected with the EDRMS adoption has been applied to the study conducted in the Ministry of Trade and Industry in Botswana. Based on the outcome of the research, the author defines efficient service delivery, transparency and cost-effectiveness, as the most vital advantages which have been achieved through the system [12]. When it comes to obstacles, the focus is made on insufficient training, similar to the previous example from New Zealand, and ill-designed user interface.

Some of the other country-specific studies discussed below are also building their hypothesis on the technology acceptance model or a combination of several paradigms.

In their study, Carter and Belanger integrated the TAM model with Diffusions of Innovation theory and web trust models to form a parsimonious yet comprehensive model of elements that greatly affect citizen adoption of e-government initiatives in the USA [44].

Sebetci's study develops and tests a (TAM)-based model of mandatory use of the Sgb.Net system by government employees in Turkish public institutions. The Ministry of Finance initiated the Sgb.Net system on June 1st 2007 to establish a measurable, analyzable, controllable and better manageable structure by transferring all financial and non-financial operations of the ministry into the electronic environment [45].

Two determinants of the TAM model - perceived usefulness and perceived ease of use has been identified as a means to predict citizens willingness to use e-government services provided by the government of Malaysia continually [46]. The core principle of this study was to figure out whether the perception of the citizens related to the consideration of using e-service is beneficial to them would lead them to keep utilising them. The study concluded that the designer must take into account the needs of users in scheming the e-government system [46].

The TPB has been adopted for studies conducted in Taiwan and Turkey. Even though in case of Taiwan, researchers Hung, Chang, and Yu investigated the case of online tax filing and payment system [47] and Ozkan & Kanat developed a model for empirical validation [48] in Turkey, both of the examples were focused on finding out determinants affecting citizens adoption of e-government services.

Various EDRMS implementation examples mentioned in this chapter reflected how different the aim of the governments might be and how it varies case by case. The common trait for each of them was that facing difficulty within the process is inevitable. Nevertheless, in most of the cases, the recommendations and lessons learned are mostly focusing on similar aspects, and they will be further elaborated and discussed in the upcoming chapters and highlighted once again in the analysing part of this study.

Table 2. Major Aspects Related to the EDRMS Implementation. Source: Author

| Challenges | Source |
| --- | --- |
| User-resistance | [41], [44], [45], [46], [47] |
| Lack of management support | [42], [43] |
| User interface | [43], [44], [46] |
| Need for cultural change | [42] |
| Lack of standardised work routines and public service descriptions | [40], [49] |

| Achieved benefits | Source |
| --- | --- |
| Improved internal records management procedure | [39], [40], [43], [45], [47], [49] |
| Improved information quality and access | [40] |
| Safe repository | [43] |
| Cost-effectiveness | [43] |
| Increased transparency and accountability | [43], [49] |
| Unified processes/determined workflows and commonly described public services | [49] |

## 2.3 Change Management Approaches

Relation of the people, systems and processes within the organization is altered significantly upon the implementation of the EDRMS [15]. When it comes to undergoing such a significant transformation, the very first stage is the change occurring at the level of an individual's mindset. They need assistance to understand the importance of shifting from paper to electronic system workflow. Therefore, change starts with people [15], and the concept of change management becomes especially relevant. "Change Management refers to the processes and associated actions and tasks required in order to manage any type of change that occurs within an organization" [15]. Organizations differ from one another, and practitioners are choosing different change management practices, respectively [50]. Nevertheless, as Adam argues, there are political, people, and system skills necessary for every approach, and they help to understand thoroughly how psychology, sociology, human and organizational behaviour play their roles in the process of change [15].

Change Management practitioners should understand organizational politics. Having political issues within the organizations, which is a collective body of teams and departments, is somehow inevitable. These issues might cause the fact that some teams and departments become resistant towards implementing a new system while others actively support it. The Change Management practitioner must grasp possible pitfalls and mitigate them accordingly. Political skills also include the idea of analytical skills since, as mentioned before, the organization consists of parts within itself, and each part holds its interests. Thus, as Adam highlights, possessing them will play a crucial role when implementing a new system; furthermore, its value remains to be vital for the evaluation of the outcome of the change [15].

An integral part of the organization is people and therefore, change management practitioners should be able to communicate and interact with them in a way that once can understand what their needs are regarding the implementation and usage of the new system [15]. It will take gathering thorough information about them and conducting proper analysis to develop the best strategy which will suit individuals, teams and departments of the organization.

As already highlighted in the discussion about the technology accepting theories, the inception of tackling resistance starts with allowing users understand why is this novelty, in this particular case shift from a traditional paper-based method into electronic, so important for their daily activities.

Altering the mindset of the people will avoid the most frequently referred "Computer Will Replace Our Jobs" syndrome [15]. While trying hard to understand the staff, those from the top of the organizational structure should also remember that "leading by example" is a compelling component. It should be first directors and leading management fully accepting and utilising the system and demonstrate amended work practices. It should be followed by the next step, which is the involvement of the whole organization, empowering staff at all levels [51].

Based on what has been mentioned above, it can be said that change is a very complicated process full of unexpected events, and it is an essential organization to be prepared for mitigating unplanned occurrences and expect unexpected events.

**Change Management Strategies**

There are two major publications from the end of the twentieth century: "General Strategies for Effecting Changes in Human Systems" (1969) by Robert Chin [52] and "The Planning of Change" by Warren G. Bennis, Kenneth D. Benne and Robert Chin [53] which have outlined three the most commonly used change management strategies. Three meta-approaches about the implementation of the change in social and organizational context are empirical-rational, normative-reeducative and power coercive. Even though these strategies have been drafted at the end of the twentieth century, they continue to be relevant and applicable to the contemporary settings of public organizations [54].

The principle of the Empirical–Rational change management strategy is that if people understand the change they are facing now will bring some benefits to them, their attitude will be positive, and they will not be resistant towards the alteration. In order for successful change to occur, it is crucial the need and the benefit of this change to be adequately communicated to them [15]. Furthermore, if there is any resistance, it should be caused by ignorance and superstition. The solution is educating individuals within the organization about the logic and the benefits of the change [51]. The critical component of the rational-empirical approach to change is information, which should be shared with the members of the organization [55], [54]. The basic principle of achieving successful application of the Empirical–Rational Change Management strategy, it is vital to be communicated with the staff members why this change is so important and what particular benefits can be obtained through this transformation.

According to the Normative–Reeducative Change Management strategy presumptions, people are striving to be accepted in the setting of their organization. The main focus of this strategy is redefining processes through collaboration [56] and finding proper associations among three core values of those of system, its members and organizational environment [54]. Even though the individuals' characteristics might include self-interest, their primary goal is beyond it and aims to achieve the formation of the human culture as they perceive themselves as equals through the input from their leader. They are not merely the recipients of the information but the participants of the dialogue which aims to achieve win-win solutions [51].

The last, Power–Coercive Change Management strategy determines that the role of authority is the most important as people will obey his commands no matter of what. Imposing external sanctions is a way to force people and achieve change. Political and economic power is used as a control mechanism to apply levels of power to those with less power [55]. Needless to say that the power-coercive strategy usually evokes anger, resistance, and damage to the fundamental relationships of those involved in the change [51]. If we consider this approach in the framework of the EDRMS implementation, it can be assumed that employees will not have much choice, they will be obliged to use the system as demanded by the leadership.

As mentioned already, every organization is different, and the strategy of change should tailor accordingly. Therefore, merely copying and pasting one of the above three approaches will not guarantee the successful implementation; hybrid approach should be developed instead [54]. As described by Adam, the most effective approach to use in public organizations with a high level of bureaucracy and resistance is the combination of the Power–Coercive and Empirical–Rational Change Management strategies [15].

# 3. Overview of the EDRMS

The following chapter aims to provide an overview of the EDRMS and look at them from three different perspectives. The first section briefly summarizes the history of its development, technical characteristics and basic functionality features. The second section discusses tangible and intangible benefits related to its implementation. The section also lists those important stages organizations should go through. The third section focuses on concepts offered by scholars which took the importance of them so high that books have been published addressing each of them separately. Lastly, in the sub-chapter, significant emphasis is put on the Estonian case, which is far ahead from other states in terms of advancement in e-governance development level. The representative of the Estonian Ministry of Economic Affairs and Communications gave a thorough insight of how EDRMS was implemented in the given ministry, what barriers they were facing, what is the current agenda and possible future shortcomings. All of the matters mentioned in this chapter will further be discussed in the last part of the research in which analysis will be made, and conclusions derived.

**EDRMS Functionality**

Before moving to discuss technical aspects of the EDRMS, it is essential to clarify its definition as it often is confused with EDMS and ERMS. Electronic Documents Management System (EDMS) is an automated system supporting the creation, usage and maintenance of the electronic documents to improve an organization's workflow [57]. Electronic Records Management System (ERMS) is also an automated system supporting the creation, usage, maintenance and disposal of electronically created records, not the documents. However, the main goal of the ERMS is to provide evidence [57]. EDRMS is the system, which incorporates both of the previously defined functionalities: document management and record management [58], [10].

The inception of recording information for future evidence has been practised by drawing pictures on the cave walls by our ancestors [15]. The act itself is very primitive; however, it can still be considered as a first occurrence of the concept of recordkeeping. Further development of the document management and record-keeping was advancing during the last century decade by decade. In 1980, pre-requisite of the EDRMS was Document Image Processing (DIP) systems. Even though they had the element of simple workflow through which scanned documents could be rotated among employees, three significant features of the system included document scanning, indexing and storing. EDMS and ERMS started first appearing in the 1990s. Even though there was no record-keeping standard, advancing the previous DIP and workflow functionality, these systems have developed into managing electronic documents and records. In 1999 the first version of the functional requirements was published by United Kingdom's National Archives (TNA). During the next couple of years, the mandatory functionality requirements for the EDRMS has started to establish. There are few common examples such as the International Organization for Standardization (ISO) 15489 issued in 2001, United States Department of Defence (DoD) updated and issued standard 5015.2-STD the following year. European standard known as MOREQ was also developed by the IDABC (Interoperable Delivery of European eGovernment Services to public Administrations, Businesses, and Citizens) [57].

Even though every organization is different and the needs they have determined the EDRMS functionality they would require, there is a list of essential elements, which every EDRMS should compy in order to be considered as a fluent operating solution [15]. The functionality follows the logical life-circle of the document from its creation until respective disposal. Together with the primary requirement of following the standards and rules (such as legislation and DoD, ISO or MAREQ standards), functionality elements can be sorted into four main categories: creation - which encompasses the components, such as repository, folder structure, integration with desktop applications, check-in and check-out, necessary for its operation; access and usage – focusing on security, accessibility consisting of version control and auditing; collaboration and workflow – which means that system has inbuilt business process management (BPM) allowing the flow of information and possibility of employees to work on the same document and exchange data; and lastly – storage, transfer and disposal, which is responsible for the retention and following disposal once the limit expires [15]. No matter what specific organizational necessities are, these aspects aligned in three categories should always be included in order EDRMS to meet its functional requirements.

**Benefits of implementing an EDRMS**

Implementation of the EDRMS solution brings tangible as well as intangible benefits to the organisations. Tangible benefits are gains directly affecting the money saved by the organisation while intangible benefits are the opposite, and they are not quantifiable in terms of monetary gain. Both types of benefits are important and worth discussing.

There are three major areas with tangible benefits: saving cost, saving floor space and productivity gains. Tasks held on paper in the filing rooms and cabinets, are now shifted into electronic methods, and it means that all the monetary resources which would have been spent on them, is saved. Moreover, the space of the filing rooms is no longer needed, which allows the organisation either to use the same space for other purposes or simply avoid the need of renting extra space. The costs saved through this might be significant to the organisation's budget. Another critical aspect of working on documents and records electronically is that organization's staff is enabled to access them instantly from their PCs instead of locating them physically in the filing cabinets. It increases efficiency in a way that staff is capable of dealing with the higher amount of work, processes faster and thus, increases productivity [50], [57]. Advanced efficiency is allowing organizations to be cost-effective and shrink the expenses, which otherwise would have been spent on staff as well as all the utilities related to working with the traditional paper-based method.

One of the primary intangible benefits that come as a by-product of successful EDRMS implementation is the accessibility of stored electronic documents and records through the centralized location via a central server [57]. It also improves staff morale. The fact that information is stored at the central location also facilitates its management process. Real-time disposition of the information allows the organization to have better control over it and improve its overall customer service by these advantaged. As already mentioned in the EDRMS Functionality Sub-chapter above, an integral part of the EDRMS is its aspect of Collaboration and BPM. Therefore, the implementation of the EDRMS can also improve the efficiency of the organization's business processes [50], [58] and with the incorporation of the workflow, documents and records will no longer be transferred manually across the organization. The fact of records and documents stored in the central location in networked storage secures the information in case of disaster – allows a full disaster to restore and recover. It has been highlighted throughout this chapter multiple times already that working with the traditional method means saving all the data in physical forms on paper which makes it vulnerable towards the natural disasters which occurrence is beyond human control. In such a scenario, all the information kept on papers will be destroyed without prospects of its restoration. This situation can easily be avoided through the backup routine of the EDRMS with the "secure off-site locations" as a mean to guarantee the disaster recovery [15].

From the list and brief description of the tangible as well as intangible benefits, it is possible to see that they are interconnected to each other. For example, if the organization can increase its staff efficiency by delivering the services faster, then it will also be positively reflected on productivity by saving costs and also, improved customer satisfaction.

Albeit the list of tangible and integrable benefits seems very appealing, the cost of all the major aspects related to the EDRMS implementation might cause resistance of the organizations. The list of the essential elements includes: "Project Management, Information Gathering and Analysis, The Feasibility Study, The Business Case, The Functional Requirements, The Technical Specification, Procuring an EDRMS software solution, IT hardware costs, Implementation costs, Training costs, Support costs, Maintenance costs" [15]. Decisive moments when implementing EDRMS within the organization is the management to understand that the initial costs at the early stage are inevitable and financial conditions will start improving only after a couple of years when there is an emergence of Return on Investment (ROI) [15]. This time varies from organization to organization, but the shorter the ROI, the better it is for them to start EDRMs adoption.

**Further insight suggested by scholars**

- *Links of the records management and knowledge mobilisation*

The connection between records management and knowledge mobilization are two essential concepts of the book by Harris 2012. The author highlights that records management triggers the knowledge mobilization in two distinctive ways. Firstly, it gives clarity to the existing situation and makes it ready for action and secondly, identifying and filling the current gaps and achieving the mobilization of knowledge as the process of discovering new knowledge, in other words knowing what is yet unknown. Records management and knowledge mobilization are complementing each other through dealing with the knowledge structure, evidence, questions of accountability and attempt to increase modernization and transformational change [59].

- *Hindering factors*

Possibility of the EDRMS implementation is also thoroughly discussed in the book by Runskill & Demb. Even though they give a very positive evaluation to EDRMS - "Databases designed to support the creation, management and delivery of electronic content, documents and records" they still list few major obstacles which might hinder their implementation in some public entities. The list consists of over-complexity, cost, suitable for large-scale operations and the need for rigid corporate record-keeping rules [60].

- *Importance of organisational culture*

Oliver attempts to explain the concept of organisational culture and why it is necessary for information managers to understand this. He brings the scenario of implementing EDRMS in four different organizational models. The first three models with the unfavourable environment for the EDMRS are marketplace bureaucracy, full bureaucracy and family model. In the case of the marketplace, bureaucracy managers rely on personal experience and record-keeping will not be viewed as essential. An entire bureaucracy in problems of widespread dissemination or unauthorized access to information is extensive. For the family model support from the leader is necessary. Workflow bureaucracy or well-oiled machine is the most suitable for EDRMS implementation [61].

- *Tools for the project managers*

Blokdyk's book can be used as a self-assessment tool for the project managers and will guide them through several primary criteria: recognize, define, measure, analyze, improve, control and sustain. They will be able to monitor and control their working groups as well as hold a proper relationship with the stakeholders [62].

The experience of implementing EDRMS in various public sector organizations is shared by Adam, who also offers direction and guidance for the public sector Managers. Through this book, they will become aware of all the motivational factors encouraging governmental institutions to shift their method of working with the traditional method into electronic. One of the highlights of the book is a thorough overview of the emerging challenges which occur together with the enhanced efficiency and productivity [15].

## 3.1 Estonian Experience of EDRMS Implementation

It was valuable for this study to have the opportunity to interview the Head of the Document Management Department from the Estonian Ministry of Economic Affairs and Communications and an independent Consultant. The aspects representative of the Estonian Ministry paid attention to was different compared to the insights received from Georgia. Nevertheless, there were still similarities which have been experienced in both country's settings. Estonian consultant's viewpoints were derived from the country's experience; however, as he emphasized during the interview, these points can apply to Georgia as well. Details will be further discussed below.

First EDRMS was introduced in Estonia in 1990, and an actual usage started a year later in 2000. Subsequently, the Public Information Act was launched, and the requirement to have all the documents and records standardised became a requirement. By the time the Act was launched, Estonian public entities were using different systems with X-Road in place. As the interviewee highlighted: *"I worked at an agency, where all the work was done in another information system, and data were exchanged, and it was very convenient."* New requirements established by the Act cause difficulties as employees have been working in the different systems, and they had to shift to the new one. The process was intensive, and employees even had to work over hours and as interviewee added: *"As record's manager told me it was a painful work because it took time to migrate everything that was needed."* Unfortunately, as later turned out, the data agency was supposed to publish sensitive data, which should have been publicly disclosed. Therefore, all this work was in vain.

In order to facilitate the implementation, excessive training took place during which staff members were introduced to how to work with the system, which retention regulations to follow, what are the principles of data privacy, who has the right to access a specific type of data and so force. In terms of the technical training, there were records managers as well as IT specialists, both supervising and making sure that both approaches were taken into consideration as they were complimenting one another. Interviewee added: *"So I think it will be useful to combine the training and have both sides explaining why and how and what are the options."* Something that remains still a bit of a challenge to the employees, as the interviewee added and which requires particular attention is the Usability and User-experience (UX): *"The problem with our EDRMS is that they are not easy to use for common workers even for me, who know a lot about it."*.

There are three important factors highlighted by the interviewees, which require further attention. First one is the excessive amount of the systems. Something the Ministries aim to achieve is having two-three systems only, which would cover every public entity without fewer interoperability issues. This idea has been announced and tried out in 2015. Nevertheless, it was difficult to measure all existing EDRMS, and thus, the process was postponed. The second topic which could be perceived as a future shortcoming is archiving, which can be split into two parts – archiving documents and archiving machine-readable data. This difficulty arises from the fact that when the system is developed the concept of the future archival need is somehow left out. Thus, this drawback will have to be dealt with in the future. As the interviewee highlighted: *"The problem with machine-readable data is in other systems, and it is true that when creating these systems, nobody thought how to archive them and how long to preserve it."* Furthermore, the project the Ministry is currently working on is the task of resource-optimisation. The inception of this idea came from the following: *"This is the main idea, and the purpose of it is that the number of records increasing rapidly. For now, when I was comparing the number of the records to the previous year of 2018 and 2019 in 41 government institutions, the number of records increased by 700,000. So the number is a bit more than 3 million, it was 2.8 before, but now it is increasing."* The study discovered that due to the small number of records in some agencies, there is no need to keep the position of the records managers. Thus, even though reducing these positions will be unpleasant, the interviewee explains that: *"We have to think what is good for the state as a whole and start understanding it might be more useful and better for resource-optimizing."*

As a summary of this Chapter, it can be highlighted that at first glance, incorporating EDRMS in the organisation's routine seems very attractive. However, as discussed in each of the above Sub-chapters, the composition of the process is relatively more complicated as people with less experience might have imagined. Although achieved benefits overweight all the struggle public entity might encounter, which is depicted in the case of Estonia, the path to this level is still challenging no matter how technologically advanced State might be. The entire passage of the EDRMS implementation in the governmental entity will be further elaborated below on the Georgian example; the case of Estonia will linger throughout the analysis and comparisons will be made.

# 4. Case of Georgia

This chapter is divided into two sub-chapters and aims to provide an overview of the case. First sub-chapter describes how Georgia got to the current e-Governance level, how the legislative framework was established for facilitating it, how service provision has shifted from traditional to digital means and its reflection on the citizens and public sector institutions.

Second sub-chapter focuses on solely the case of implementing EDRMS in the governmental institutions. The discussion started with the description of establishing document management standards by the end of the twentieth century when only paper-based documentation was in use. It also describes how the initiation of using EDRMS arose and what legislative and technical measures were taken.

## 4.1 Overview of Georgia's e-Governance Story

Georgia, like many other countries, is attempting to develop e-Governance at national as well as local government levels. In order to grasp the inception of the current EDRMS challenges, adding context by mentioning a few details from the country's recent history will be helpful. By the year 2006, Georgia – a country with 70 years of Soviet legacy, was left with empty treasury as well as the extremely centralized and overregulated economy. Problems were especially severe at the level of governmental organizations. Nepotism, corruption and bureaucracy were an integral part of the workspaces, which lead to the unskilled workforce and lack of motivation among employees. As a result, there were either poor governmental services or no services at all. Until 2006 there were no registries, and everything was on paper [63].

Already from 2006, Georgia started making small yet steady steps towards the improvement as a State. It also reflected on its e-Governance conditions. The most essential and initial step was creating primary registries and core data centres. The Government of Georgia declared the development of e-government services as one of the priorities in 2009 and started to implement different projects. For example, the Ministry of Justice established two Agencies during 2010-2013 for supporting the development of the public electronic services - Data Exchange Agency and Public Service Development Agency. As enablers of the e-service development, Government adopted new laws: Law on Creation of Data Exchange Agency (2010) [64], Law on Unified Information Registry (2011) [65], [66], Law on Information Security (2012) [67], Law on Personal Data Protection (2012) [68] and e-Document and e-Signature Law (2017) [69]. Few examples of the successful service platforms developed during the last ten years are as follows: e-declaration www.rs.ge – electronic portal to submit tax and customs declarations, applications and complaints to the Revenue Service and also do money transfers; Unified system of procurement, e-tenders www.procurement.gov.ge - allows electronic participation in various governmental tenders, advertising requirements for goods and services, registering customers and suppliers as well as issuing and receiving relevant documents; Citizen Portal www.my.gov.ge – allows access to the information and services unified in different categories: family, healthcare, social services, assets, business, online payments.

Georgia was one of the first countries which joined OGP in September 2011. By making this step, the country expressed its willingness to commit the basic principles of the partnership: transparency, accountability, citizen participation and technology & initiative through which country would conduct governance reforms and improve public services through implementing innovative technologies. Until now, there were four action plans developed: 2012-2013 [70], 2014-2015 [71], [72], 2016-2017 [73] and last, National Action Plan of Georgia 2018-2019 [74], which has been developed by the Ministry of Justice of Georgia in cooperation with civil society, international partners and citizens. The country has made progress during the past eight years. Few the most significant achievements include the availability of the public information electronically, establishing electronic petition portal - www.ichange.gov.ge through which citizens can submit petitions to the Government of Georgia, creation of very first open data portal www.data.gov.ge and one-stop-shop for 64 public services available online at citizen portal (www.my.gov.ge) [75].

## 4.2 EDRMS in Georgia – Background and Description

In 1999, the President of Georgia issued an edict on establishing general rules and guidelines for managing documents within Georgian public institutions. The edict obliged, Georgian State Chancellery, Council of Ministers of the Achara and Abkhazia regions, other governmental agencies, as well as local authorities to organize the implementation of standard procedures in their central offices and as well as in the subordinated offices. The fulfilment of this responsibility required them to develop document management regulations and instructions within three months after the edict entered into force. By the end of 2000, they developed exemplary nomenclatures for the affairs of homogeneous institutions, organizations and enterprises and, in agreement with the State Archives Department of Georgia and introduced them in these institutions. The 53-page long Unified Rules of Procedure is a normative document describing what the standard of each document is and is legally binding for every Georgian public institution [76].

Georgian Civil Service Bureau's research of Georgian public institutions in 2011 showed that the Ministry of Finance, Ministry of Internal Affairs, Ministry of Justice and the Ministry of Defence have been using sophisticated and technologically advanced documents management systems [77], [78]. Nevertheless, these systems were not interconnected, experienced interoperability issues, and thus, there was no document exchange between them. The Bureau aimed to facilitate the implementation of the documents management systems in every public institution and accomplish this goal by the beginning of 2012. Bureau prepared and presented the project of electronic document management system's minimal standards with the active participation of the Ministry of Finance of Georgia, the Ministry of Internal Affairs of Georgia, the Ministry of Justice of Georgia and the Ministry of Defence of Georgia (programmers), as well as employees of the LEPL Data Exchange Agency. After the launch of this program, all types of correspondence, including citizens' statements, would be transmitted electronically through the Ministry's system, greatly simplifying the access of documents to the addressee, ease processing and responding to them. The main goal of the electronic document management system's standard was to adjust existing programs to a unified minimal standard, as well as to consider these requirements when designing future programs. It would significantly reduce expenses in public institutions and save time and human resources. For developing requirements, thorough studying and analyzing of the best international and Georgian experiences took place.

As a conclusion of all these activities, the Civil Service Bureau developed a strategy for their unified Document Management System in 2011. The strategy examined disadvantages of managing paper documents and overviewed existing circumstances, in particular, developmental stages of the EDRMS, statistical data about their implementation in different public entities, and how were the processes developed. In the second part, the focus was on strategic directions, interoperability standards among different systems, data exchange scheme and priorities as well as integral components for the development of the project.

Subsequently, in 2012, the Government of Georgia issued a decree on approving a minimum standard for document management systems in public institutions [78]. The outcome of the decree would be reduced time and resources spent on managing documents manually, make processes more rationalized and sophisticated based on the statistical data generated by the system, standardise procedures across the central and local governmental organizations. Shift to the electronic system would ease the coordination between the institutions with remote physical locations and ensure effective communication within the staff members of the particular organizations. As mentioned earlier, by the time the decree was issued, some of the public institutions were already using EDRMS. Those who were not using them yet had the plans to implement them in the nearest future. However, there are two main problems, which are highlighted in the decree and tackling them is the primary goal of it: lack of efficiency and economy of those systems when developed locally and the imperfections of existing local systems and their incompatibility. Public institutions are trying by themselves to develop a system tailored to their needs locally; designing and creating an algorithm for the same product consumes a large number of funds. One of the most significant drawbacks here is that there are not many intellectual and financial resources to develop a high-quality product; thus, the goal remains unachieved, and in the end, they get an imperfect and incompatible local system. Usually, they lack complete information regarding the document life-cycle, and quite often, some integral features and capabilities are also missing, which makes it unable to exploit the system with its full advantages. Implementation of the unified standards would facilitate the development of quality and effective solutions and guarantee the compatibility of programs implemented in different institutions.

The third Article of the Government's decree mentioned above also established the list of the conditions documents management systems should fulfil, which is the following:

All modules of the program should be interconnected and exercise uninterrupted exchange of information. At the same time, the modules must be autonomous and able to operate outside of the unified system. Each module's capabilities and functions should be based and derived from the existing legislation. The adjustments should be possible to occur in case of a change in the relevant normative context.

The need for manually entering information should be minimal through presenting all possible fields as unified classifiers and linking to existing electronic databases. The system should also respond automatically to detected errors and should not be allowing saving changes which conflict with the algorithm embedded in the system. At the same time, in case of change, the date and author of the action must be recorded automatically. The system should provide access to primary data for those individuals who had access rights. Access rights should be strictly defined, monitored and granted. The system should be compatible with other automated systems already implemented at the intermediate level (for example, with the human resource management system, and so force). It should be able to upload and do various operations on the electronic files of different formats. The documents from the system should be printable and have the legal power, which is being regulated by a special Act. All functions should be designed based on intuitive principles and thus, convenient and comfortable for beginners to understand. Considering all these requirements, 233 public institutions now employ one out of three information management systems created by either the Ministry of Internal Affairs (named "e-FLOW"), Ministry of Justice (named "DES") or Ministry of Finance (named "eDocument").

For example, by the end of 2016, the National Agency of Public Registry of Georgia has implemented it in 78 organizations. The core principle of this system is to improve the official correspondence and document exchange between public organizations, manage tasks efficiently and reduce administrative burden. The system is considered to be a facilitator in the development of e-Governance. It allows users not only to communicate with other entities, manage their documents, use the flexible search engine, save time and resources, use in-built MS Word and MS Excel but also sign a document digitally. Currently, there are more than 245 public and private organizations and more than 20,000 end-users.

A year later, an amendment in the 2nd Article determined which Ministries were responsible for delivering the program. By 2012 only the National Agency of Public Registry (Ministry of Justice of Georgia), Analytical Financial Agency (Ministry of Finance of Georgia) and Education Management Information Systems (Ministry of Education and Science of Georgia) were able to provide software which corresponded to established minimum standards [79]. There was another amendment in 2018, and the list of the Ministries changed as following: the National Agency of Public Registry (Ministry of Justice of Georgia), Analytical Financial Agency (Ministry of Finance of Georgia) and Georgian Operational-Technical Agency (State Security Service of Georgia) [80].

## 5. Interview and Questionnaire Outcome

Empirical data provided in this chapter has been collected through the interviews to the Georgian and Estonian representatives. For the thorough investigation of the case of EDRMS implementation in Georgia, respondents at three different levels were selected. The Minister of Education and Science was chosen as a respondent since he was actively involved in the transformation from the paper-based to the electronic document management systems. His input is useful for understanding the mindset at the state level when the implementation was decided, and a decree issued by the Government. His, as minister's incentives, were different from the other two respondents, and it will further be discussed below. On the other hand, the Head of the Document Management Department of the Ministry of Culture and Monument Protection of Georgia gave a more detailed overview of the implementation procedures. Since she was responsible for choosing the software, making desirable adjustments in the system based on the Ministry's requirements and was always helping her employees, interviewing her provided the data which otherwise would be challenging to collect. Therefore, if the Minister was able to give a more holistic view of why EDRMS was needed, what was the main goal and strategy, the person, who was actively engaged in the implementation process, provided more practical input. The importance of understanding the context when doing research has already been highlighted in the chapter addressing the research design and methodology. Therefore, it was perceived as crucial to investigate further and understand the context of the EDRMS implementation at the local government level. The person who was interviewed was Governor of the Lagodekhi Municipality and currently, holds the position of the Chairman of the City Council. Interviewing him was beneficial to understand how the processes were directed at this lower level and what measures were taken.

## 5.1 Insights from the Initial Stage

As mentioned before, the implementation of the EDRMS on the state level became mandatory in 2012 when the Government's decree was published. Even though the implementation as we will see later, did not go without barriers, overall assessment of the process received from the Minister of the Education and Science in 2009-2012 was positive: *"The system itself was quite easy. There were three major factors, needed for its successful implementation: network, which works correctly, good training module and the will. When all these three components were on place, it was easy to implement."*

An in-depth overview of how this policy was realised was shared by the representative of the Ministry of Culture. According to the interviewee, it was the Administrative Department together with the Document Management Division under her supervision directly involved in this process. There were three programmes provided by the Ministry of Justice, the Ministry of Internal Affairs and Ministry of Finance, from which public institutions were able to choose the most desirable ones: *"I studied all three programs, then summarized their pros and cons (based on our activities and the specifics of the Ministry) and presented findings to the deputy Minister"*. Another essential detail the representative of the Ministry of Culture explained was working in the Test Mode, during which they were carrying out their duties on paper, as usual, while also testing accomplishing tasks through the system. She added: *"It helped us to identify flaws while building a database, whether we provided incorrect information, or whether the document should move differently. It was the best way to learn more about this program simultaneously. After switching to live mode, we had fewer mistakes and fewer problems."*

One of the difficulties highlighted by the representative of the Ministry of Culture, which was also the case in Estonian example, was enormous work they had to go through when switching between the systems; and the Ministry of Culture has been reorganized multiple times. Nevertheless, as the witness and participant of each transformation, she explained that the first significant change when the two Ministries – Ministry of Sport and Ministry of Culture joined, was trouble-free to overcome since both of them were working in the same system ("DES"). However, it was not the case with the second expansion when the Ministry of Education came onto the board, and they were using the different programme – "eFlow".

**Role of the Management**

As the Minister mentioned during the interview: *"One of the most critical factors of the successful implementation was that there was a will to do it."* Undoubtedly, government's inclination towards the successful implementation of the ICT is the crucial factor, nevertheless, the attitudes of the people working on the level of the institutions who are working day by day to make things happen, is as much necessary. Responses received from the other two respondents prove that as well. The representative of the Ministry of Culture mentioned: *"I personally always welcome these innovations and new possibilities"* and that in 2010 when she started working as a Head of the Document Management Department when everything was registered on paper, she started creating simple charts on Excel spreadsheet to find documents quickly. Another initiative their department raised was the need for future archiving. It would have been a mutual project between the Ministry of Culture and the National Archive of Georgia. Nevertheless, their initiative failed due to the unification with the Ministry of Sport. One more aspect which became evident from the interview was the willingness of the Managers of being involved in the process themselves. The representative of the Ministry and the Chairman highlighted their participation in everyday activities from the very first stages of implementation.

**Goals of the public entities**

The direct correlation of the EDRMS implementation with the improved service delivery has been emphasized by the Minster and the Chairman. According to the Minister: *"It was especially crucial since our Government's main goal was to reduce citizens' communication with the bureaucracy as much as possible and made maximally automated."* He also provided an example of the project between the Ministry and the Public Service Hall (Legal Entity of Public Law of the Ministry of Justice). This strategy worked out, and Public Service Hall issued certificates of secondary education together with many other documents instead of the students collecting them from the schools. *"Our goal was to establish one-stop-shot for the citizens in the Public Service Hall."* - he added.

Efficient communication with public entities and improved service delivery as the primary goal has also been highlighted by the Chairmen of the Lagodekhi City Council. He commented: *"it is our primary purpose and the reason why people chose us. We can respond to their requests in a shorter time. It is quite essential, as the citizen does not have to wait two or three weeks for an answer."*

**Staff training**

The significance of staff training has been highlighted by all the respondents. As the Minister of Education commented: *"Staff training started at the very first stage of the implementation process, and the Government monitored it."* The system was in test mode in the Ministry, which lasted for up to three months. He said: *"It was followed by additional training just to make sure all of the essential topics have been covered, and no additional explanation was needed."* Nevertheless, as the representative of the Ministry of Culture added: *"Even though the test mode was quite long and employees had theoretical knowledge, shifting to the normal mode was still more or less painful. However, if it continued like this, it would go on forever, this test mode and this training, and there would always be that fear, and we would never be able to overcome the fear of working in this normal mode."*

The situation was a bit different in case of the Lagodekhi Municipality. Since there were not many people who had training, it was the responsibility of one person to introduce the system to the rest of the City Council employees. As the Chairman explained, it was a demanding task for this person as there was extra work he was supposed to accomplish and get paid as usual: *"I was fortunate that the appointed person in charge was a person who was not lazy and was receiving the same salary and was doing extra work to teach these employees and start a job."* Nevertheless, he conducted his duty successfully and even though staff members still needed assistance during the next couple of months, the main goal was achieved, and they shifted their performance to the digital method.

## 5.2 Identified Barriers

**Adjustment to the New System**

One of the staff-related challenges brought out by all the respondents, except for the Minister, was the adjustment to the new way of working. As the Minister explained: *"The Ministry of Education and Science was using EMIS (Electronic Management Information System) since 2009 and due to the experience of working in this system, adopting EDRMS was relatively easy for the majority of the staff. We can say that they were more prepared with this pre-requisite compared to other employees of different Ministries."* It most certainly was an issue for the Lagodekhi Municipality as the Chairman commented: *"For most employees, this was a new way of working (e.g., how to write a letter to an employee and other technical details), and for about two to three months, that person in charge helped as needed."* The situation was similar for the Ministry of Culture as: *"One of the first was the adjustment of the current document management rule to this electronic program. Various details should have been spelt out more clearly in the program."* In addition to that, as she later elaborated, they had to readjust the whole staff tree and make sure that the new system was in exact alignment with their organisational structure. This concept has also been highlighted in the interview with the Estonian Consultant, where he emphasised that *"Usually they (staff) need assistance with the different functions or features of the system."*

The representative of the Estonian Ministry of Economic Affairs and Communications also commented that: *"The problem with our EDRMS is that they are not easy to use for common workers even for me, who know a lot about it… they are not convenient to use." … "They (e.i. staff) have to know why this is necessary, but also they have to experience that something is easier now than before."*

**Perception as a Separate Artefact and the life-circle of the document**

Estonian expert also accentuated two factors which are very important in the initial stage of the implementation and will cause shortcomings in the future if proper attention is not paid. As he said: *"One of the biggest challenges I have encountered when it comes to the implementation of the EDRMS is that they are being treated as a separate artefact."* He clarified this by further adding that usually it is the administrative department within the organization or Ministry which is responsible for the implementation and it could be somehow isolated and not be the integrated part of the whole organization. The second one, which is closely related to this: *"When it comes to the implementation of the EDRMS whole life-circle of the document, should be taken into consideration."* Therefore, it should be well-thought ahead how the system will be managed, how the workflow will work, what the business-process is like and also, how the archiving will work. It is essential to understand that there are not only text-based formats composing the EDRMS but also another more considerable amount of data which is being stored in different machine-readable formats. Usually, they are easier to process and archive, nevertheless, they need it to happen should be pre-determined and well understood.

Furthermore, another advice shared from the Estonian experience is the extreme necessity of the standardization from the early stage. As the Representative of the Ministry of Economic Affairs and Communications elaborated: *"I would emphasize the high importance of standardization of all document and record types and processes."*

**Resistance to Change**

Two of the interviewees highlighted the resistance to change: *"But we have to bear in mind that people get used to even to a bad system. Thus, when they have worked with a not very good system, for a long time even if the new one is a bit better, they are not always happy. It is very common with every system, I think."* (Representative of the Estonian Ministry) and *"The older generation protested to some extent that we have been working like this for so many years, and now this is what you are inventing, what are you going to achieve with that."* (Chairman of the Lagodekhi City Council).

When there is an expected change in the organization, the very first thing people become concerned and scared about is the risk of losing their jobs. Such fear has a disturbing effect over peoples' minds and pushes them to see only negative pictures. It is easy to assume that under such circumstances, employees of the organization would instead try to resist the change as much as they can instead of facing the fear and jeopardize their current "safe" environments. As the Chairman of the Lagodekhi Municipality recalled: *"They were afraid that someone else might be able to do the job better than them; they might be reprimanded and lose their job."* The resistance is intensified even more when people do not have much information about the upcoming change and what exactly should they expect. Fear of the unknown drives them to combat whatever is going to happen and either prevent it from happening or alienate from it as much as they can.

**Fear of Failure**

Fear of failure can be described as a barrier composed of all the obstacles mentioned above. It also urges people to feel hesitant to incorporate the new system in their working environment. As the representative of the Ministry of Culture described the implementation process: *"Despite the test mode, the day of transition to live mode and not just the day, the whole month, I can say, was pretty hard. Staff members were afraid and sceptical. Their concerns were about storing documents, preserving their work, and also how they were performing specific tasks. So, all of this first phase, the first month and quite a long time straight away from me and my coworkers, who went through long training, and worked harder on ourselves and the program to be more prepared and mobilized for those first days when we went into real mode. We could no longer do our job, and the priority was helping our staff. We went to the rooms on the spot and helped them with report cards, announcements, or letters. We were actively involved at the beginning stage. We had to work until late because at the same time we had to do our direct work and it was quite tricky, I can say, and this period was hard."* Even though using the EDRMS is mandatory, there are specific procedures, which are voluntary to be accomplished through it. In such cases, people are allowed to prepare a hard copy of the document, sign it manually and use the organization's stamp, however, they are still requested to register the document electronically, which means that they have to scan the document and initiate the registration procedure. Therefore, there is a duplication of work. It was to save peoples' time and resources and facilitate workflow, but the result is the opposite. One more factor, which is hindering the usage of the system and leads to the manual work, is the fear of the system errors. People tend to choose the method which might require more time and effort from them if it means that they will not mess something up in the system and feel incompetent. As the Chairman of the Lagodekhi Municipality mentioned: *"To some extent, I explained this by the fact that they found it challenging to work electronically and were less able to master these computer technologies."*

Another "negative effect" which has been named by the representative of the Ministry of Culture, based on her experience was employees complaining about the lack of human interaction. It could be a country and culture-specific as some other countries least interaction with people is even more favourable. She commented on the following: *"The only negative thing we all consider is less communication with employees. Before that, our doors were always open, documents were brought* *in for redirection or registration, and we were always in contact with all the employees of the Ministry, and we were a little bored."*

## 5.3 Achieved Benefits

After describing the main reasons of the EDRMS implementation and identified barriers, interviewed government official also highlighted all the benefits the system brought to their organizations. They can be divided into two major categories, as described in one of the previous chapters: tangible and intangible benefits.

A very first tangible benefit which seemed most noticeable to the interviewees was the reduced amount of paper and thus, saved cost on that: *"first and foremost, it saves paper" (Chairman of the Lagodekhi Municipality City Council).*
Saving human resources has been one of the significant benefits highlighted by the representatives of Georgian and Estonian Ministries, and each of them brought examples. The Ministry of Culture and Monument Protection of Georgia has many other agencies and organizations which are subordinate LEPL (Legal Entity of Public Law). For delivering documentation to them, there was a designated position of the courier providing this service across the city. When there was a need for them to deliver the document urgently, there was a need for fast delivery. For that, they also had a car and a driver who was supposed to accompany the courier. Therefore, together with reducing the cost of the employees now, the cost of gasoline is also saved. She elaborated: *"After the introduction of this program, we abolished the position of the courier. That is, human resources are saved: the driver, the courier plus fuel costs, and time."*

The aspects of reducing staff costs have been highlighted by the representative of the Estonian Ministry of Economic Affairs and Communications. Nevertheless, the emphasis was put on the position of the records manager. As she explained, the number of records in different governmental agencies varies and is usually based on their size. They have encountered cases when there are not many records created in the EDRMS. Thus, the position of the records managers is gradually being reduced.

Another important component the Minister of Education and Science paid attention to was achieved time-efficiency: *"The most important was achieved time-efficiency. The amount of time needed for searching for and managing documents has significantly reduced."* The time factor has also been emphasized by the representative of the Ministry of Culture. She mentioned how much time it would take them to send a particular document to another agency or Ministry and said that: *"Now, the letter automatically delivers in the dedicated agency right after signing, and it takes seconds literally from the signing to the delivery of the document."*

The importance of EDRMS at the local government level was depicted from the Lagodekhi Municipality City Council Chairman. He explained briefly how time and resource-consuming communication with the central government has been: *"We used to send documents to the Ministries by post; it took days to get there, additional time to reach the minister or deputy; but now it is a matter of minutes, is quick and efficient."*

He also pointed out how much the coordination with the other local government institutions had become: *"Before the implementation of the EDRMS, we had to send physical copies of the documents to the Lagodekhi City Hall; it was time-consuming. Additional time was needed before the record would reach the Mayor and then some more before we received a response from them. Now, this is very much simplified as both of the entities work in the same program (we wall it doc-cycling). In my opinion, it is beneficial not only at the municipal level, but it also facilitated communication with the ministries and various agencies, who also have EDRMS."*

Removing physical barriers and convenience has also facilitated the speed of the employees' performance, as the Minister emphasized: *"It was also a great support to the departments since they no longer needed to search for different documents in the office rooms as staff members had access to the databases, based on their positions. It was very convenient for them."*

The challenge of shifting to the electronic system increased workflow efficiency. Chairman of the City Council explained: *"We had to build a system. We had to hold a few meetings about it, and we talked to the deputies and the heads of the service about it."* He explained that before the implementation of the EDRMS, there was a designated space (called Service Bureau) in the Governor's office, where the employees were sitting and receiving citizens' inquiries. They were collecting citizen letters in folders and putting them all together in his office. After that, he had to assign them using their organizational scheme. Usually, it would take a couple of days to reach the responsible personnel: *"This, of course, affected the service, the relationship with the citizens; they may have needed a quick response…we to have a much more sophisticated service now."*

Having clarity in terms of the staff tree and the workflow, interviewees emphasized an increased sense of responsibility among employees. The Minister pointed out that: *"Because there is electronic management, the sense of responsibility from the staff members is higher; they are no* *longer able to hide and quietly fix their mistakes. It increases their commitment to do the task properly."*

A similar viewpoint was mentioned by the Chairman of the Lagodekhi City Council: *"Through the system, I knew who was assigned to the particular task, whether the deadlines were met or not...It had become easier to control them."* The opportunity of traceability and evidence has also been highlighted by the representative of the Estonian Ministry: *"Some benefits are that everything is traceable, you can see who did what and when, so that it can be used in court."* and the Chairman as well: *"This process of assigning each other was also documented, these documents were then stored, and if something went wrong, we could take out the records and see that it was not only the responsibility of the Governor but also of all these participants."*

## 5.4 Understanding Users

The design of this study allows looking at this single-case from two different angles. In the previous chapter, the perspective of the state has been analysed. In this chapter, the attention will be shifted towards the employees. Therefore, as the second unit of analysis, civil servants of various Georgian public institutions have been inquired.

In order to gain information and support the matters stated in the Research Question 2, the survey has been drafted and distributed among Georgian public servants through social network and other electronic channels of communication. An overall number of received within a month is 101 (from mid-March till mid-April 2020) which aimed to describe employees' attitude towards the usage of EDRMS in their daily activities. Overview of the questionnaire outcomes in the first sub-section of the present chapter is followed by their analysis in order to answer the imposed research question fully. Furthermore, lastly, at the end of this chapter, risks and limitations for the given research method will be discussed. The interview and questionnaire outcome from this chapter will contribute to conclusions and recommendations at the end of this study.

The conducted survey consisted of 10 questions and aimed to inquire the EDRMS features employees perceive to be the most beneficial based on the characteristics of their work, whether they had system training if they need assistance in terms of accomplishing their tasks and overall attitude towards the usage of the EDRMS. The survey was anonymous, and participants' personal information has not been collected. Descriptive statistics of each of the question is listed as an appendix to this research.

The goal of the initial first two questions of the survey was to collect information of the public institution participants were employed in and which EDRMS they were working with. Based on the answer, the list consisted of the 8 Ministries, 11 LEPL (Legal Entity of Public Law), 4 City Halls and 2 City Councils, as well as 11 other governmental institutions (See Appendix 3, Table 3 for the complete list). As for the EDRMS used, there were three of them prevailing eFlow with 36.6% (37 respondents), eDocument with 28.7% (29 respondents) and DES - 15.8% (16 respondents). Remaining 18.8% (19 respondents) replied that there were using different systems from the mentioned three, which were: DSSI (Used in the Lagodekhi Municipality City Hall), MSDA (Tbilisi City Council and City Hall) and ERP (LEPL The State Military Scientific-Technical Center "Delta") (See Appendix 3, Figure 2).

As mentioned in the theoretical framework, the importance of system training has been frequently highlighted as a critical factor when it comes to technology acceptance among staff members. The third question of the survey was aiming to investigate this matter. As it turned out, 31.7% (32 respondents) have not had any system training, 39.6% (40 respondents) replied that they had training before they started working on their current position and 28.7% (29 respondents) had training before they started using the system and later on as well (See Appendix 3, Figure 3).

In the fourth question, participants were asked whether they need assistance for accomplishing their daily tasks through the system. As the polled civil servants responded, for 39.6% (40 respondents), there is no difficulty for solving their problems and no additional assistance from their supervisor or colleague is needed. Slightly more than the half of the participants 52.5% (53 respondents) polled that they need assistance, but it happens rarely, and only 7.9% (9 respondents) need help frequently (See Appendix 3, Figure 4).

In the fifth and sixth questions, participants were allowed to write their own opinions and answer the questions based on their personal experience and specificity of the tasks. In the firth question, respondents were asked to indicate which features of the system they perceived to be the most useful. Answers received from them were different; nevertheless, it is possible to see several unique patters. The first three most frequently referred preference is related to time-efficiency, control mechanism and improved communication. The following, sixth question, which is the logical continuation of the previous one, aimed to find out which daily tasks were they accomplishing the most frequently through the system. Answers varied here as well, which is natural as it is based on their positions. Nevertheless, the following tasks seemed to be similar to most of the respondents: writing reports, sharing/forwarding/assigning tasks to coworkers and engaging in the correspondence within the organization and with external public entities.

The seventh question asked citizens what an average percentage of the daily work is they accomplish per day. The possible slots included four different levels: 0-30%, 30-50%, 50-70% and almost 100%. According to the responses, 17.8% (18 respondents) chose first option of 0-30%, for 22.8% (23 respondents) it was second option of 30-50%, 27.7% (28 respondents) opted for third one – 50-70% and lastly, 31.7% (32 respondents) are accomplishing almost 100% of their task using EDRMS (See Appendix 3, Figure 5).

Following the eighth question focused on whether public sector employees would choose to accomplish a specific task on paper rather than in EDRMS if they were given the freedom to choose. Majority of the responses were negative, and 85.1 (86 respondents) would not use the paper-based method. For 10.9% (11 respondents), there are cases during which using paper might be more convenient rather than the electronic system. Only 4% (4 respondents) poled that yes, they would instead use the old method (See Appendix 3, Figure 6).

The ninth question investigated the age of the participants. Answers to choose from were divided into four categories: 18-25, 26-35, 36-50 and more than 50. According to the data provided, 17.8% (18 respondents) were within 18-25 age category, 39.6% (40 respondents) in the second of 25-36, 33.7% (34 respondents) in the third category of 36-50 and 8.9% (9 respondents) were more than 50 years old (See Appendix 3, Figure 7).

The last, tenth question focused on the experience of using EDRMS. It was less than a year for 15.8% (16 respondents). Almost half of the respondents, consisting 47.5% (48 respondents) replied that they had one to five years' experience and 36.6% (37 respondents) have more than five years' experience (See Appendix 3, Figure 8).

Forthcoming sub-section of this chapter will focus on conveying analyses of presented questionnaire outcomes and discussing matters related to sample representativeness as well as outcome generalizability.

## 5.5 Analysis and Discussion

Analysing outcomes of this study lead to the below-presented interpretations, which help to give an overview of the EDRMS implementation in Georgian public sector. Received responses from the interview participants as well as answers to the survey questions were crucial in terms of developing findings to the initially stipulated research questions. Even though the implementation of the EDRMS was mandatory for the public entities, empirical research showed that the concepts discussed in the ICT acceptance theories discussed above are still applicable.

The personal attitudes and perceptions highlighted in the Theory of Reasoned Action and Diffusion of Innovation as an initial barrier, were also mentioned by the interviewees. Those three interviewees who were actively engaged with the staff members during the process of implementation (representatives of the Ministry of Culture of Georgia, Ministry of Economic Affairs and Communications of Estonia and Lagodekhi City Council) have provided a detailed overview of the resistance they have faced. In the case of the Ministry of Culture, scepticism from the staff members caused the resistance as they expressed distrust towards the technology. In the case of the municipality, a major focus was resistance to change. Dent (1999) stipulates that systems of social roles, with their associated patterns of attitudes, expectations, and behaviour norms, share tendencies to resist change, to restore the previous state after a disturbance [81]. Thus, it is in human nature to be inclined to stability. In order to look at this challenge from the third - employee's perspective, two variables of system training and system's daily usage have been correlated using Pearson's Correlation. The outcomes revealed that there is a positive correlation and employees who had training are incorporating EDRMS in their daily activities compared to those, who replied as they did not have any system training (See Appendix 3, Figure 9). Here comes the relevance of the Technology Acceptance Theory as well. As Davis argues, in order for people to use technology two, the most significant criteria should be fulfilled: they should perceive that system is useful and that it is easy to use. Proper training can be used as a facilitator for achieving this result. However, as mentioned by the interviewees, in case of Lagodekhi Municipality, there was only one person who had proper training in the capital city of Tbilisi and then he had to train every single employee back in Lagodekhi and be in their assistance when the support was needed. The situation was slightly better in the case of the Ministry of Culture as there was a department responsible for the implementation. However, a couple of people, including the head of the Document Management Department, were actively involved in training and assisting their staff members. While sharing Estonian experience, representatives of the Ministry of Economic Affairs and Communications confirmed that there were Records Managers and IT personnel mutually delivering training, which can be considered as a most feasible solution to the proper delivery.

The outcome of the public servant survey also led to another correlation which is between the working experience, daily usage of the system and needed system assistance. Research showed that the longer an employee has worked in the EDRMS, the higher percentage of using the system for accomplishing daily tasks is. Thus, Pearson Correlation is positive. Correlation is negative in terms of the second variable - needed system assistance, which means that the employees with longer EDRMS working experience need the least help from their managers or coworkers. Here the Unified Theory of Acceptance and Use of Technology is applicable according to which this is one of the significant determinants (See Appendix 3, Figures 10 and 11).

Interview, as well as survey respondents, listed some of the EDRMS essential features and functions, which are widely used in their everyday activities. The most frequently referred elements were repository, version control, search and retrieval, as well as auditing. However, there are several more integral elements which stay beyond proper attention. Something interviewees discussing Estonian case wanted to point out was the essentiality of standardization and clarity of future archiving measures. These two factors have not been much discussed by Georgian governmental officials. It was only the representative of the Ministry of Culture who said that it was her initiative to bring up the case of archiving and urge the Ministry to raise the topic. Nevertheless, it never succeeded. It is not too long since Georgian government institutions shifted their traditional method of working into electronic management. In this particular case study Ministry of Culture started form 2012 and Lagodekhi City Council from 2014; however, thinking through the retention and disposal of the records, which determines the length of time records should be kept and disposing of them when the limit is up, is essential.

As mentioned above, EDRMS implementation became mandatory in 2012 when the Government of Georgia issued a decree and established minimum standards or document management, which should have been met by the public institutions. However, because organizations are using systems offered by different software vendors, there is an obstacle to system incompatibility. Even though the element of collaboration is met and people and teams within the organization can communicate and share information, for instance, to work on documents together, when it comes to sharing data with the other institutions they are facing difficulties. Thus, the urgent need for taking into consideration the standardisation of the documents is essential.

Interviews reviled that organizational change has occurred in all three discussed cases. Ministry of Culture of Georgia, Ministry of Economic Affairs and Communications of Estonia as well as Lagodekhi Municipality City Council had to undergo significant changes in order to adjust this new way of working. Such a big change is not only a challenge itself but also greatly affects other aspects of the implementation and might lead to significant barriers. Therefore, even though the concept of leadership has not been mentioned by the interviewees, the notions discussed by them confirms its importance and presence in each of the provided examples. As discussed in the UTAUT Theory, managers must understand what the drivers of the technology acceptance among employees are and what leads to subsequent successful usage [35], [82]. Active participation of the Chairman as well as the representative of the Ministry of Culture and the Minister himself can be considered as one of the facilitating factors. Furthermore, as the Chairman of Lagodekhi City Council described, it was him the initialling implementation of the EDRMS in the office and in order to ease the process of adoption he was personally conducting a meeting in order to help employees understand the positive effect and necessity of this change.

## 5.6 Limitations of Interviews and Questionnaire

Yin highlights logical tests, such as validity and reliability, as the most useful assessment tools of such exploratory case study [24]. Constructing validity helps to identify correct operational measures of the studying object. It consists of two parts – internal and external. The goal of the internal validity is to establish a causal relationship through which the others trigger certain conditions. External validity displays whether and in what context the findings of the case study can be generalized. Also, reliability shows that data collection procedures will have the same results when repeated. Fulfilment of these criteria has been attempted by investigating the context and using multiple sources of data. All the conclusions drawn from them have been made through their simultaneous analysis. Nevertheless, the validity of this research will probably be jeopardized if the same study will take place in the future. The reason for this will be the fact that the internal settings, as well as external factors, will have significant influence and might alter the results and not be possible to imitate.

Another important aspect, which requires particular attention and can be considered as an essential risk to validity is the bias. Researchers pick specific topics due to their interest, and as Willis mentions, in such cases, they already are expecting what the results and conclusions might be. Therefore, it is essential to understand this instead of pretending to be objective [22]. The fact that the researcher is personally collecting data can also cause selective observation. The author of this thesis has experience of working in one of the public organizations in the named state and is aware of the study subject. In order to avoid developing the bias and reduce it as much as possible, appropriate preventive measures were taken. Such measures include avoiding researching the public organization where the author of this thesis was employed. Guidelines of the semi-structured interviews were followed, and they were conducted in a neutral tone, recorded and created detailed transcript without personal interpretations. The ideas of the respondents were indicated separately so that they could have been easily recognised by the reader and not be perceived as the author's perspective.

This research aimed to interview all of the stakeholders to investigate their viewpoints. As mentioned earlier, LEPL Financial-Analytical Agency of the Ministry of Finance is the software vendor in quite many public institutions. The representative had agreed on a face to face interview in February 2020, when the author planned visit to Georgia. Nevertheless, due to the force-major situation caused by the Covid-19 outbreak, the Government of Georgia declared a state emergency and travel was no longer possible. The representatives of the Agency have not agreed on conducting an online interview.

Existing pandemic circumstances have greatly affected the collecting of survey respondents. Public organisations have not been responsive in terms of sharing questionnaires to their employees, and the author managed to collect 100 responses by sharing them individually with the public servants.

# 6. Conclusion

Presented research revealed that even though national governments are inclined to implement ICT in their activities in the attempt to improve their internal processes and service delivery, the process of change is not effortless and they have to endeavour significant challenges.

In response to the first research question, which was posed within the frames of this research, the main EDRMS implementation challenges were analysed. Before moving directly to the Georgian case, an overview of the ICT adoption theories, as well as country-specific EDRMS implementation examples, were discussed. Particular attention was paid to the case of Estonia, which stands among the most advanced e-states with highly efficient digital governance. Estonian experts also shared their experience and described an existing situation. For this exploratory study, Georgian government officials were interviewed who gave detailed information on the implementation process and the challenges they were facing. Based on the results of this inquiry, the major obstacles included: technical difficulties, especially at the initial stage when the first system was launched; organisation-wide confusion which meant perceiving EDRMS as a separate artefact and not fully incorporated within the entity and last but not - the human factor.

Furthermore, as cited the most in the literature, resistance to change and fear was also prevailing among Georgian public servants. The study clearly showed that it included factors of losing a job, increased accountability and lack of IT literacy.

After depicting all the significant hindrances, the focus was shifted, and the second research question aimed to explore what measures were taken by the Georgian government entities to combat against them. The research results demonstrated that the role of leadership was tremendous. Those who were directly involved in this transformation became personally responsible for achieving desired outcomes; hence, the fact that the process of change was somewhat successful can significantly be assessed as a result of their dedication.

In order to answer the third question and identify benefits achieved through the usage of EDRMS, together with the related literature, Estonian case study and interviews, actual users were inquired. The survey among employees of various Georgian public institutions gave a different perspective on how staff members perceive this shift from traditional to the paperless method. The examination identified that attained advantages include saving paper, saving human resources, succeeding in time-efficiency, removing physical barriers, security, transparency and an increased sense of accountability. Furthermore, factors which were the most commonly described by the employees were improved communication, clarity of tasks, the efficiency of the workflow and document management, in general.

## 6.1 Prospects of the Future Work

As already mentioned in the literature review, scholars are applying different technology acceptance theories to country-specific case studies. What I have realised throughout this research is that there is no "blueprint" which will take most of the challenging factors into consideration and offer the solution for solving them. Instead, the hybrid model, which is based on a few theories and also takes into consideration the specificity of the country's context, should be the most feasible. Because there has not been such a study in a Georgian setting, going further and performing research which will be based on testing a hypothesis, will be useful. Combination of three models - TAM, DOI and UTAUT will be essential for building them. Since they focus on different aspects such as user perceptions, the role of leadership, and expectancies with moderating variables, the outcome of the study should be descriptive and reflect the situation from different angles.

# Appendix 1 – Interview Questions

Questions to:

A representative of the Ministry of Culture of Georgia – Audio recording 08.04.2020
A representative of the Estonian Ministry of Economic Affairs and Communications
Audio recording 02.04.2020
Chairman of the Lagodekhi Municipality City Council – Audio recording 16.04.2020

1. Would you please briefly describe the implementation of the EDRMS in the Ministry of --- (according to the respondent) (when and how did it start, which program are you using, what was the strategy, responsible departments, etc.)?
2. What main challenges did you encounter during the implementation?
3. How would you describe the experience of using EDRMS compared to a paper-based method (what are the benefits or downfalls, if any)?
4. How did the staff members reacted to this change and shift to EDRMS?
5. What measures were taken to facilitate the implementation and usage of the EDRMS by the staff?
6. Would you please describe the communication with one of the public entities through EDRMS (it could be an example of a specific e-service or a project you were together working on).

Questions to the Minister of the Education and Science of Georgia 2009-2012
Audio recording 14.03.2020

1. When did the Ministry o Education and Science start implementing ESRMS?
2. Did the Ministry develop its implementation strategy?
3. What were the main challenges in the process of the implementation of EDRMS?
4. When did the training of the Ministry's staff take place?
5. How often was the training conducted?
6. What was relatively older staff members' attitude towards the usage of the new system?

7. What was the procedure or the challenge concerning the system and accomplishing tasks, which seemed to be difficult for the staff?
8. Was there research or questionnaire investigating the dependence and attitude of the staff about newly implemented EDRMS?
9. Which benefits, in particular, would you name which were brought by the usage of EDRMS compared to the traditional, paper-based method?
10. When did the Ministry start cooperating with other Ministries through the EDRMS?
11. Were there any interoperability issues identified within the process of cooperation with other ministries?

Questions to the Consulting Manager at PWC Advisors – Audio recording 01.04.2020

1. How the implementation strategy of the EDRMS is usually designed and which major factors are being taken into consideration?
2. In which areas do the challenges/obstacles appear when it comes to implementation of EDRMS in the public organisations?
3. Based on your experience, what are the ways to tackle those challenges mentioned in the previous question?
4. How would you assess the importance of Change Management when it comes to the implementation of the EDRMS in the public organisation?
5. What is the most commonly requested "support" or "help" from the public entities? / In which areas do they usually struggle the most and require further assistance?
6. What is the (most common) feedback/response you receive from the public entities after the implementation of the program?

# Appendix 2 – Mindmap of Interview Outcome

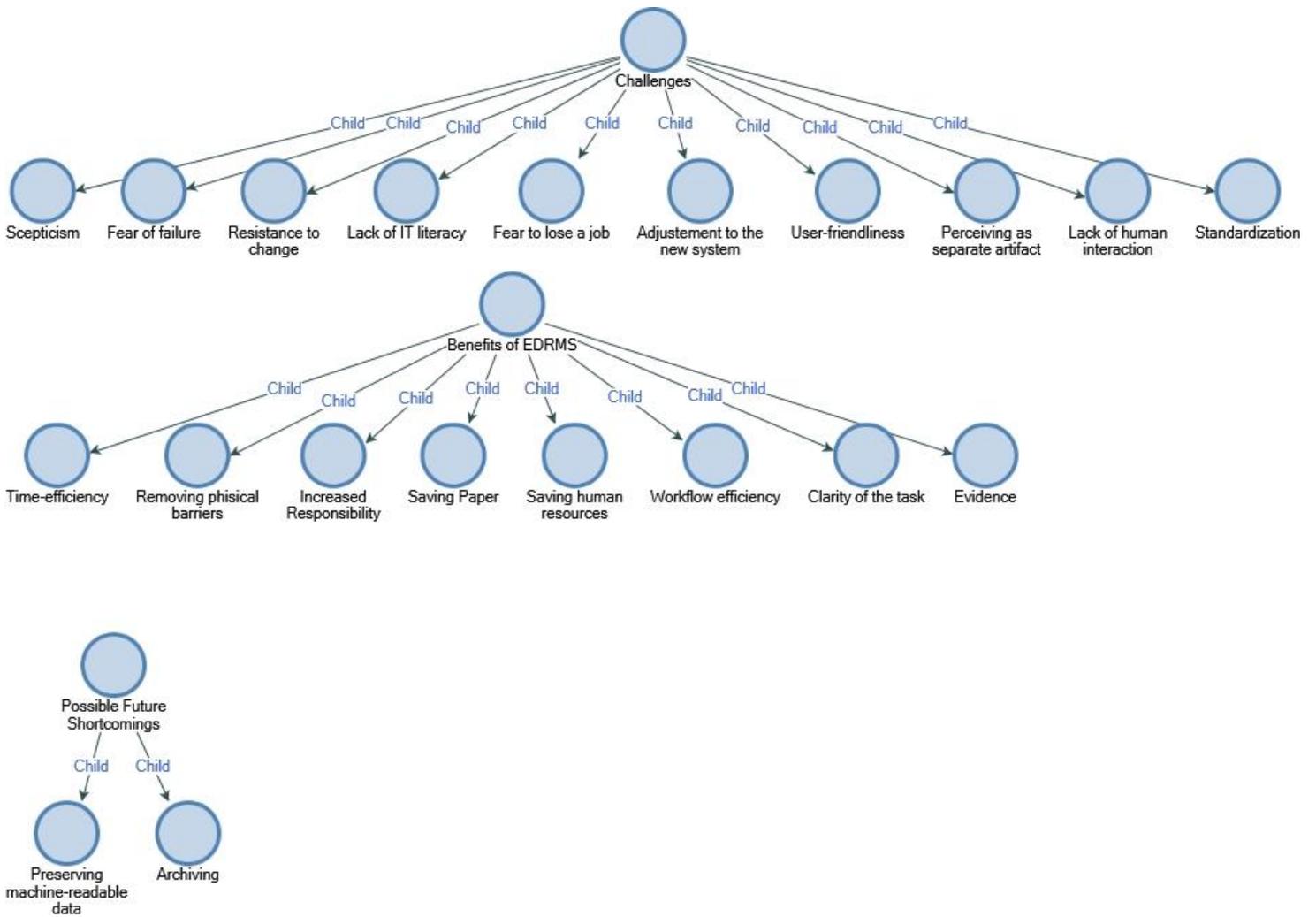

Figure 1. Key Concepts from the interviews

# Appendix 3 – Results of Questionnaire

This questionnaire was created to assess public servants' attitudes towards EDRMS usage in their everyday activities. Participants were recruited by posting a survey on social media platforms, contacting directly and other means of pair to pair digital networking. Data collection lasted for a month (from mid-March until mid-April 2020). Analysis of the outcomes has been presented in Chapter 5 of this study. Below are listed descriptive statistics of the results.

## Q.1 In which public institution are you working?

| |
|---|
| Georgian Chamber of Commerce and Industry |
| Georgian Government Administration |
| Gurjaani Municipality City Hall |
| Lagodekhi Municipality City Hall |
| LEPL Agency of Protected Areas |
| LEPL Emergency Response Center "112" |
| LEPL Enterprise Georgia |
| LEPL National Defence Academy |
| LEPL of the Ministry of Education, Science, Culture and Sport of Georgia |
| LEPL Office of Resource Officers of Educational Institutions |
| LEPL Operative-technical Agency of Georgia |
| LEPL Shota Meskhia Zugdidi State University |
| LEPL Social Service Agency |
| LEPL Tbilisi Opera and Ballet Professional State Theatre |
| LEPL The State Military Scientific-Technical Center "Delta" |
| Ministry of Defence of Georgia |
| Ministry of Economy and Sustainable Development of Georgia |
| Ministry of Education, Science, Culture and Sport of Georgia |
| Ministry of Environment and Natural Resources Protection of Georgia |
| Ministry of Finance of Georgia |
| Ministry of Foreign Affairs of Georgia |
| Ministry of Internal Affairs of Georgia |
| Ministry of Justice of Georgia |
| Municipal Development Fund of Georgia |
| National Agency of Georgian Public Registry |
| National Statistics Office of Georgia |
| Parliament of Georgia |
| Revenue Service of Georgia |
| Rustavi City Hall |
| Rustavi Municipality Service Center |
| Special Penitentiary Service |
| State Institute of Literature |
| State Security Service of Georgia |
| Tbilisi City Council |
| Tbilisi City Hall |
| Tbilisi State University |

Table 3. Answers to the Question 1

## Q2. What is the name of the EDRMS that you are using?

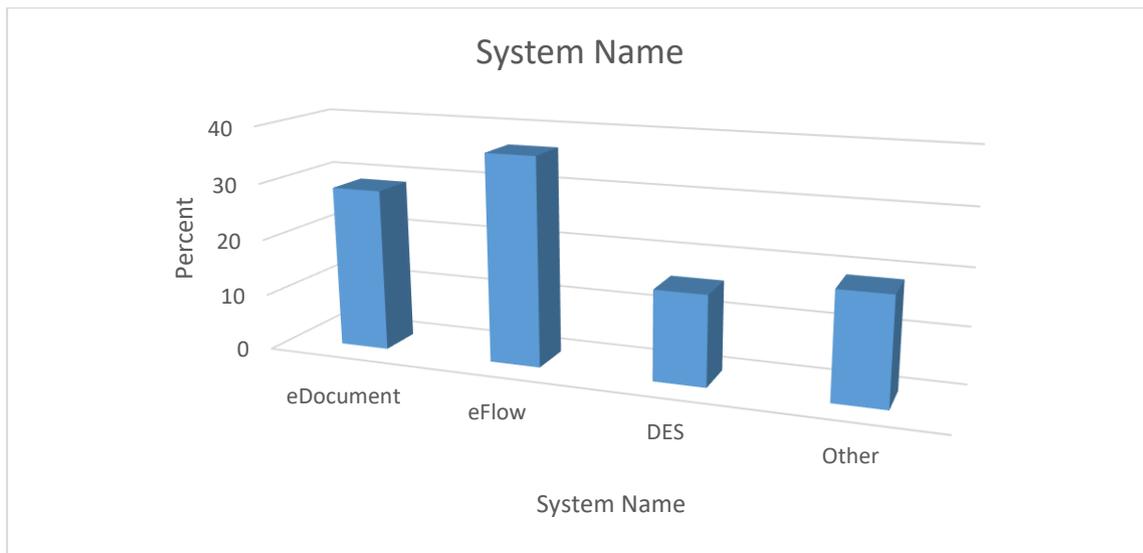

Figure 2. Answers to Question 2

## Q3. Did you have the training of the system before start using it or after?

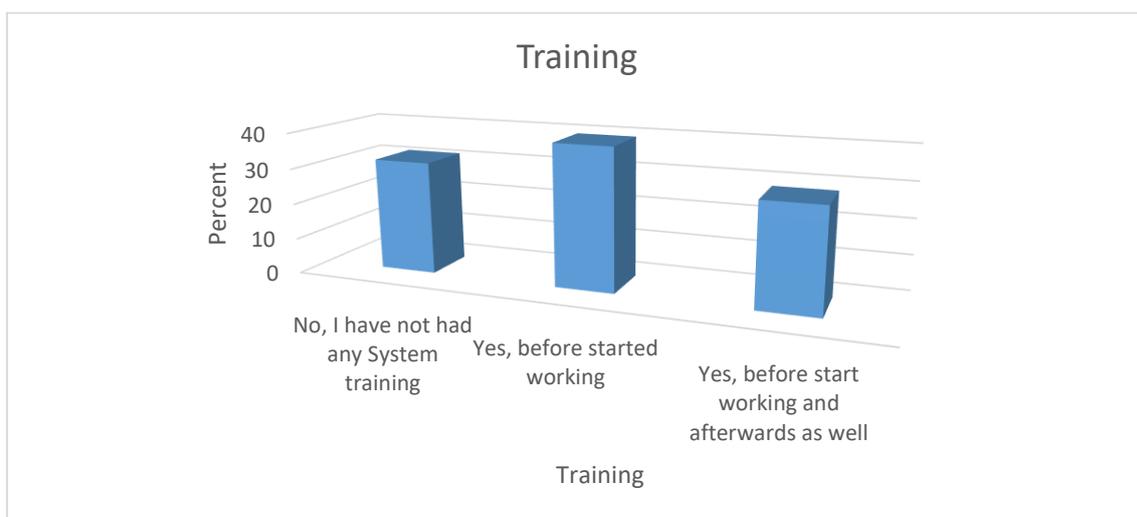

Figure 3. Answers to Question 3

## Q4. How often do you need support from your manager or coworker to fulfil specific EDRMS command/request?

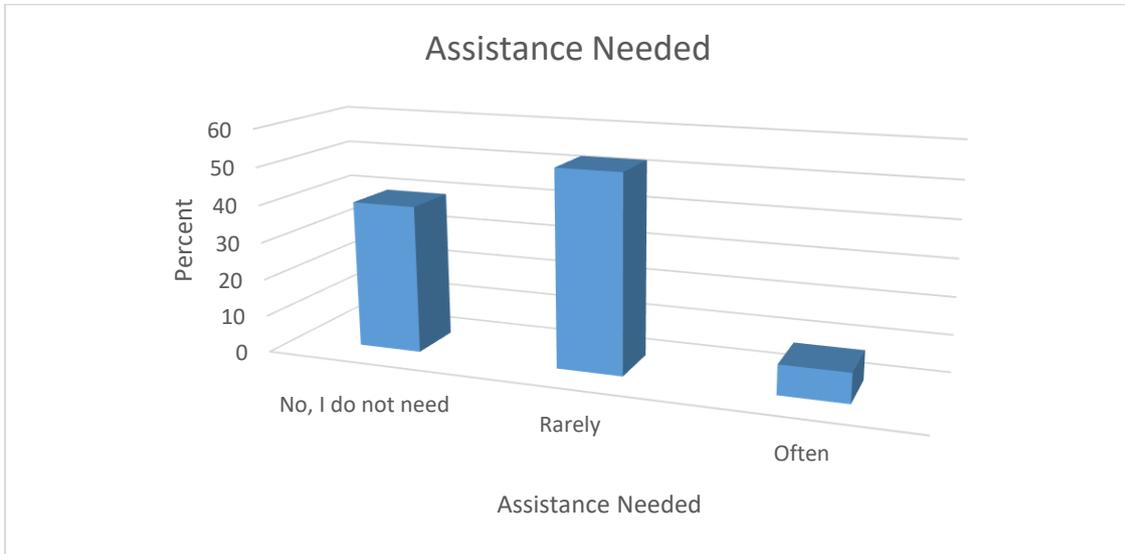

Figure 4. Answers to Question 4

## Q7. What is the average percentage of the daily work you do through EDRMS?

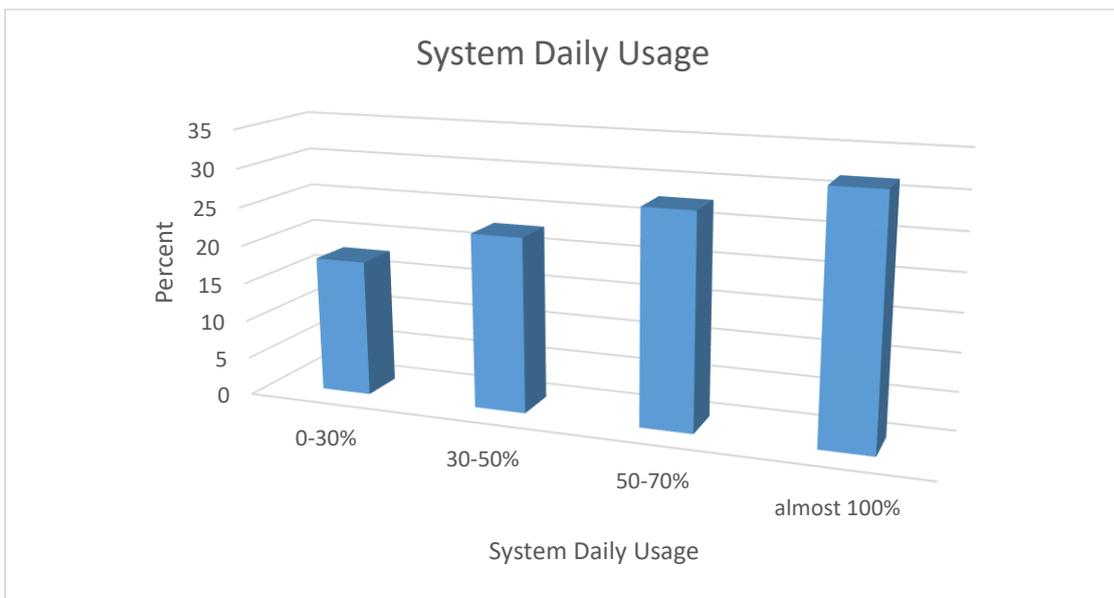

Figure 5. Answers to Question 7

**Q8. If you had the freedom to choose between accomplishing your task via paper or EDRMS, would you choose to work on paper?**

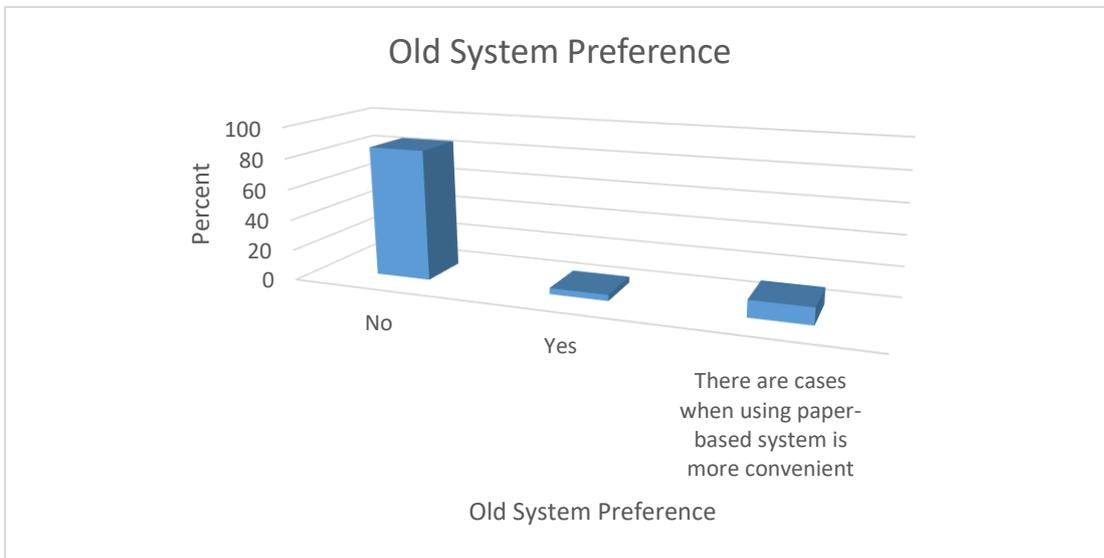

Figure 6. Answers to Question 8

**Q9. Please indicate your age**

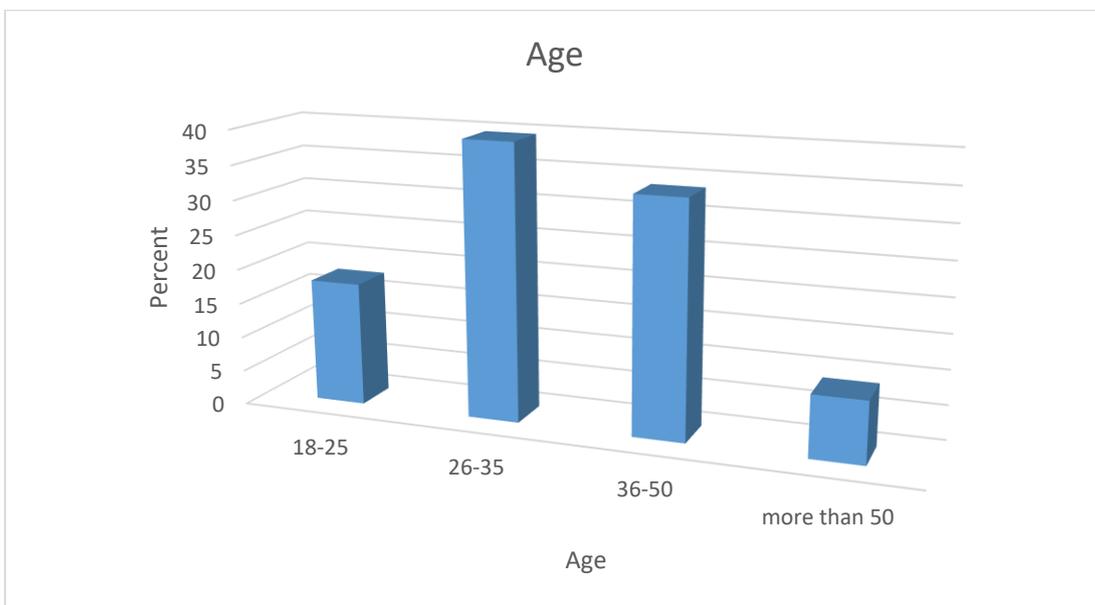

Figure 7. Answers to Question 9

## Q.10 Please indicate your experience of using EDMRS

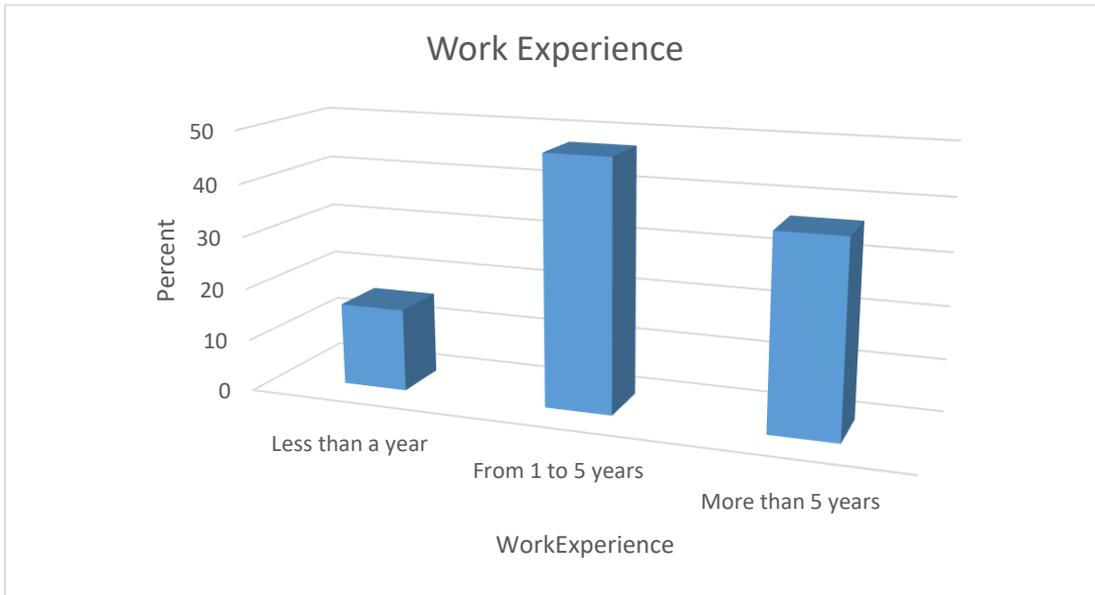

Figure 8. Answers to Question 10

## Pearson Correlations

|  |  | Training | System Daily Usage |
|---|---|---|---|
| Training | Pearson Correlation | 1 | .295** |
|  | Sig. (2-tailed) |  | 0.003 |
|  | N | 101 | 101 |
| System Daily Usage | Pearson Correlation | .295** | 1 |
|  | Sig. (2-tailed) | 0.003 |  |
|  | N | 101 | 101 |
| **. Correlation is significant at the 0.01 level (2-tailed). | | | |

Figure 9. Pearson Correlation between training system daily usage

|  |  | Work Experience | System Daily Usage |
|---|---|---|---|
| Work Experience | Pearson Correlation | 1 | .283** |
|  | Sig. (2-tailed) |  | 0.004 |
|  | N | 101 | 101 |
| System Daily Usage | Pearson Correlation | .283** | 1 |
|  | Sig. (2-tailed) | 0.004 |  |
|  | N | 101 | 101 |
| **. Correlation is significant at the 0.01 level (2-tailed). | | | |

Figure 10. Pearson Correlation between work experience and system daily usage

|  |  | Work Experience | Assistance Needed |
|---|---|---|---|
| Work Experience | Pearson Correlation | 1 | -.264** |
|  | Sig. (2-tailed) |  | 0.008 |
|  | N | 101 | 101 |
| Assistance Needed | Pearson Correlation | -.264** | 1 |
|  | Sig. (2-tailed) | 0.008 |  |
|  | N | 101 | 101 |
| **. Correlation is significant at the 0.01 level (2-tailed). | | | |

Figure 11. Pearson Correlation between work experience and system assistance needed